# Ultrasound-Assisted Facile Nanoconfinement of Luminescent Guests in Zirconium Metal-Organic Frameworks

Yiyi Wang,[1] James D. Taylor,[2] and Jin-Chong Tan[1,*]

[1]Multifunctional Materials & Composites (MMC) Laboratory, Department of Engineering Science, University of Oxford, Parks Road, Oxford OX1 3PJ, U.K.
[2]ISIS Neutron and Muon Source, Science and Technology Facility Council, UKRI, Rutherford Appleton Laboratory, Chilton, Didcot OX11 0QX, United Kingdom.

[*]*Corresponding author: jin-chong.tan@eng.ox.ac.uk*

An accelerated sonochemical route for fabricating luminescent guest–MOF (LG@MOF) composites is demonstrated, for which light-emitting molecules are nanoconfined within porous zirconium-based hosts using probe ultrasonication rather than conventional bath sonication. Ultrasonication enables the rapid synthesis of crystalline UiO-66 and UiO-67 nanocrystals within 10 minutes, with material yields of up to 92%, and facile *in situ* guest encapsulation within 1 hour, hence notably faster than the traditional 24-hour solvothermal route. Using UiO-66 and the larger-pore UiO-67 as porous hosts, tuneable single- and dual-guest fluorescent host-guest systems incorporating Rhodamine B (RB) and Fluorescein (F) were achieved. Photophysical characterisation showed broadband emission, tuneable colours, high quantum yield and improved lifetime, highlighting the synergy between sonochemical conditions, structural defects, and photophysical behaviour. Additionally, the sonochemically prepared LG@UiO particles show well-defined crystal facets and potential as photoactive thin films for solid-state lighting, demonstrating their promise as a tuneable light source for optical sensing and nanophotonics.

# 1 Introduction

Improved lighting techniques that outperform the conventional ones in both efficiencies [1] and properties are continuously being sought. For over a century, artificial lighting relied on inefficient thermal emission or plasma-based discharge processes. The transition from blackbody radiation to bandgap-engineered electroluminescence marked a fundamental shift in photon generation. Solid-state lighting (SSL), enabled by electron-hole pair recombination in semiconductors, now dominates high-efficiency illumination. Likewise, improvements in spectral quality, photostability, tuneability and materials sustainability drive new advances in luminescent down-conversion materials for SSL [2].

Despite the widespread applications of electroluminescence in solid-state lighting, the potential of photoluminescence for SSL remains underexploited. In photoluminescent lighting, ultraviolet (UV) excitation sources or blue light emitting diodes (LEDs) induce down-conversion emission from a photoactive material layer (often comprising inorganic phosphors) [3], where the efficiency and applicability are governed by photoluminescence quantum yield (PLQY), excited-state lifetime, and resistance to UV, thermal and photochemical degradation [4, 5].

Metal-organic frameworks (MOFs) are versatile inorganic-organic hybrid materials featuring tuneable porosity and high internal surface area [6], enabling potential applications ranging from gas storage and separation, catalysis, to sensing and drug delivery. Unlike LEDs and organic (O)LEDs, which emit light directly, the luminescent guest–MOF (LG@MOF) composites developed in this study are designed as UV-excited spectral converters with emphasis on colour tuneability and photostability. However, isolated molecular fluorophores are susceptible to aggregation-caused quenching (ACQ), vibrational non-radiative relaxation, and environmental instability under prolonged irradiation or elevated temperatures [7]. Confining photoemissive guests within porous MOF hosts has therefore emerged as an effective strategy for enhancing excited-state integrity and photostability [8].

The Zr-based MOF, termed UiO-66, is constructed from 12-coordinated nodes of the $Zr_6O_4(OH)_4$ clusters, conferring better chemical and thermal stability compared to many other MOFs [9, 10]. Their modular structure comprises dicarboxylate linkers with variable length and functionality while retaining the fcu topology, resulting in isoreticular analogues such as UiO-67 and UiO-68 [11]. Henceforth, UiO-66 type MOFs were selected as the host materials in this study given their structural robustness and intrinsic linker fluorescence, which enhance blue emission difficult to achieve using guest dye molecules alone. The larger rigid $\pi$-conjugated structure of biphenyl 1,4-dicarboxylic acid (BPDC) in UiO-67 further boosts fluorescence by narrowing the energy gap between the highest occupied molecular orbital and the lowest unoccupied molecular orbital (HOMO–LUMO gap) and suppressing non-radiative decay pathways. Furthermore, the relatively low toxicity of Zr [12] and the reported biocompatibility of UiO-type MOFs [13] are encouraging for use in biosensing and bioimaging.

Although structural defects are often regarded as undesirable in MOFs, they can be deliberately introduced to enlarge pore apertures, reduce framework density, and strengthen guest–host interactions through defect engineering. In UiO-type frameworks, missing-linker defects are more readily generated and less detrimental to framework stability than missing-cluster defects, making them particularly valuable for tuning pore environments. Such defects can be introduced by manipulating coordination equilibria during nucleation and crystal growth or through post-synthetic lattice modification [14]. Solvothermal and hydrothermal synthesis achieve this through competitive coordination, linker deficiency, and control of pH, water content, or temperature [13]. Microwave-assisted synthesis promotes kinetically trapped linker vacancies by accelerating nucleation [15], whereas mechanochemical synthesis favours high defect concentrations through restricted diffusion and mechanically driven bond formation [16]. In contrast, diffusion-controlled crystallisation generally produces lower defect densities with greater spatial ordering [17]. Post-synthetic modification further enables selective generation of missing-linker and missing-cluster defects through bond cleavage or component removal [17-19].

The application of ultrasound in chemical synthesis has been investigated since the early 1980s [20]. Ultrasound powers liquid-phase reaction systems through acoustic cavitation, where microscopic bubbles rapidly expand, shrink, and implode, generating transient localised high-energy microenvironments [21] that accelerate chemical processes. In MOF synthesis, these cavitation-enhanced conditions promote rapid nucleation and kinetically trapped crystal growth, making ultrasound-assisted synthesis particularly suitable for defect engineering through the formation of missing-linker defects [22, 23]. Compared with conventional solvothermal methods, it offers faster synthesis, lower thermal requirements, smaller particle size [24], and improved control over particle morphology [25]. For LG@MOF systems specifically, replacing prolonged solvothermal heating with ultrasound-assisted synthesis can mitigate thermal degradation of fluorophores while simultaneously control defect distribution and guest encapsulation. UiO-66 and its derivatives, including UiO-67 and Zr-NDC, are often regarded as synthetically demanding MOFs, because intense heating and long reaction times are commonly required for framework formation. Therefore, developing a simplified and accelerated ultrasound-assisted route is essential for expanding the practicality of LG@MOF composites to yield photoluminescent SSL.

Luminescent guests (LGs) can be incorporated into MOFs through the *ex situ* or *in situ* pathways, differing in the sequence of host-guest structural assembly. In *ex situ* approach, fluorophores are introduced into the pre-formed MOFs, whereas *in situ* encapsulation allows the framework to assemble around dispersed guest molecules. Although post-synthetic encapsulation has been applied to dyes, drugs, and metal ions [5], it is often restricted by pore accessibility, uneven guest loading, surface adsorption, dye aggregation, and ACQ, particularly for bulky molecules. In contrast, *in situ* confinement can enhance guest-host interactions, suppress aggregation, and improve encapsulation efficiency and photostability. Rhodamine B (RB) and fluorescein (F) [26] were selected as luminescent guests because of their well-established photophysical properties, complementary emission profiles to the MOF host, and sensitivity to local chemical environments [27]. Their moderate molecular size, planar structure, and $\pi$–$\pi$ interactions favour confinement within the UiO frameworks, where rigidification suppresses non-radiative relaxation and enhances PLQY [28, 29]. The spectral separation between RB and F additionally enables dual-guest emission and potential guest–guest energy transfer studies [30].

By combining the aforementioned principles, herein we demonstrate that acoustic cavitation employing probe ultrasonication can be leveraged to accomplish rapid MOF formation, while enabling *in situ* confinement of luminescent guests through a defect engineering strategy. The advantages of ultrasound-assisted *in situ* synthesis [31] could pave the way towards LG@MOF composites with tunable photophysical and photochemical properties, targeting photoluminescent SSL, optoelectronics, and optical sensing.

## 2 Experimental Sections

### 2.1 Materials

Zirconium (IV) chloride ($ZrCl_4$), fluorescein (F), methanol ($CH_3OH$), and polyvinylpyrrolidone (PVP) were purchased from Sigma-Aldrich. Terephthalic acid (BDC) and biphenyl 4,4-dicarboxylic acid (BPDC) were purchased from Fluorochem. *N*, *N*-dimethylformamide (DMF), hydrochloric acid (HCl), and glacial acetic acid ($CH_3COOH$) were purchased from Fisher Scientific. Barium sulfate ($BaSO_4$) and rhodamine B (RB) were purchased from Acros.

Deionised (DI) water was produced in house. All the reagents were used without further treatment or purification.

## 2.2 Synthesis protocols and parameter tuning

All ultrasonication probe treatments were performed on the Fisherbrand™ 505 Sonicator equipped with a stainless-steel probe (Ø6 mm) located in an acoustic enclosure. The output power was set to 20 W, generating a displayed amplitude as 10% of its maximum capacity.

**Synthesis of UiO-66:** For the solvothermal control group, the protocol was modified from the methods reported by Katz *et al.* (2013) [32], where a 24-hour oven treatment at 120 °C was applied. In the ultrasonicated protocol, 1 mmol of $ZrCl_4$, 1 mmol (or 1.3 mmol) of BDC, were dissolved in 10 mL of DMF. After homogenisation, 2 mL of 3.7 M HCl was added to the reaction mixture, which was subsequently probe ultrasonicated for 10-30 minutes. After the reaction finished, the white precipitate was washed first with DMF and then with methanol, before activation in vacuum oven at 120 °C for at least 3 hours.

**Synthesis of UiO-67:** The ultrasonication synthesis protocol of UiO-67 was based on that of UiO-66, but by employing an optimised metal-to-linker ratio M:L ($ZrCl_4$: BPDC) of 1:1.4.

**Two-acid UiO synthesis protocol:** Instead of a single-acid reaction (2 mL HCl or 4-6 mL for $CH_3COOH$), 2 mL of 0.1 M HCl and 2 mL of glacial $CH_3COOH$ were combined in a standard reaction system composed of 20 mL DMF.

## 2.3 Materials characterisation

**Sample preparation for characterisation:** For atomic force microscope (AFM) and field-emission scanning electron microscope (FESEM) imaging, the samples were prepared *via* drop-casting of very diluted dispersions of the wet synthesis product particles in methanol onto clean silicon wafers. The drop-cast samples were air dried before imaging. Samples for SEM imaging were sputter-coated with a 10-nm thick Pd/Au layer with Quorum SC7620 sputter coater. For measurements with the spectrofluorometer, the powder samples were prepared by grinding the dried reaction products with a mortar and pestle.

**AIQ control study:** The physical mixture contrast sample was prepared through immersing the powder of a non-emissive material (e.g. $BaSO_4$) in a dye solution (with methanol) of calculated concentration and drying at 120 °C overnight, to make a physically mixed powder at the same

LG weight percentage as calculated for the corresponding LG@MOF sample from its TGA curve.

**Characterisation techniques:** AFM images were collected with an Asylum Research Cypher S AFM under air tapping mode. FESEM images were acquired using a TESCAN LYRA FIB-SEM, while powder X-ray diffraction (PXRD) measurements were performed with a Rigaku MiniFlex diffractometer equipped with a Cu $K_{\alpha}$ source (1.541 Å). Attenuated Total Reflectance Fourier-Transform Infrared (ATR-ATR-FTIR) spectra were obtained with a Nicolet iS10 spectrometer equipped with a diamond attenuated total reflectance (ATR) cell.

Solid-state fluorescence spectra (excitation and emission profiles, excitation–emission maps), luminescence quantum yield (QY), and time-correlated single photon counting (TCSPC) decay were recorded with an Edinburgh Instruments FS-5 spectrofluorometer equipped with the SC-30 and SC-10 sample modules. TCSPC measurements were performed using a 365 nm EPLED picosecond pulsed laser source, with $BaSO_4$ as the reference. Lifetime fitting and quantum yield calculations were performed using the Fluoracle software. Corrections for scattering were applied to all fluorescence measurements.

For quantum yield measurements, blank spectra were recorded using flat polytetrafluoroethylene (PTFE) plugs in place of the sample holders. Sample measurements were performed under both direct and indirect illumination, and QY values were calculated from the corresponding blank, direct, and indirect emission data. During fluorescence profile measurements, excitation and emission bandwidths were adjusted to maintain photon counts of approximately $10^5$, while a bandwidth of 17 nm was used for fluorescence decay measurements.

Photostability of the materials was evaluated by conducting accelerated photostability tests in FS-5, monitoring both emission intensity and quantum yield. Accelerated photostability test involves excitation of the tested luminescent material at 80 times the operational voltage to simulate working condition over the material's lifespan, as long as 9600 hours, within a 5-day testing period. Emission intensity was measured by continuously exciting the sample at its peak excitation wavelength and recording emission spectra at increasing time intervals up to 7200 min (5 days). To avoid interrupting the irradiated surface, quantum yield measurements were performed on another sample taken from the same batch under continuous 395 nm excitation, corresponding to the operating wavelength of the solid-state lighting device.

### 2.4 Solid-state lighting device

**Device Setup:** The proof-of-concept UV-excited solid-state lighting device is composed of a 395 nm UV-emitting diode, a black 3D-printed housing that channels the light and fixes the sample film on the focal plane for even lighting, and a portable spectrofluorometer (UPRtek MK350S) connected to a monitor equipped with uSpectrum software for real-time spectral data collection and analysis. See SI, Section 4.1 for photos of the device setup.

**Thin-film sample preparation:** Activated LG@MOF (or MOF) powder was dispersed in a saturated aqueous solution of polyvinylpyrrolidone (PVP) until completely opaque, and the mixture was drop-cast onto a thin, clean glass slide (if additional blue emission is required, PET substrates could be used instead of glass slides). The drop-casted sample was dried under air flow on a flat surface to yield an even distribution of LG@MOF nanocrystals. When the sample was completely dried, a thin glass slide was placed on top of the sample, and the edges were sealed with a double layer of Kapton tape. The thin-film samples have an average thickness of 0.5±0.1 mm.

## 3 Results and discussion

### 3.1 Structural defects in UiO-66 and UiO-67

Electron microscopy reveals a clear distinction in crystallisation behaviour between UiO-66 and UiO-67 under identical sonochemical conditions (**Fig. 1a, b**), despite the same fcu topology and space group $Fm\bar{m}$-3$m$ (**Fig. 1c, d**). UiO-66 forms aggregates of faceted nanocrystals with a relatively uniform size, whereas UiO-67 exhibits more heterogeneous and agglomerated particles. This contrast reflects the influence of linker length on nucleation and growth kinetics, where the larger BPDC linker in UiO-67 slows nucleation, increases the tendency towards the more kinetically favoured pathway, and is more sensitive to reaction conditions [10, 18].

Atomic force microscopy (AFM) further highlights these morphological differences. UiO-66 nanoparticles display relatively flat and square surfaces (**Fig. 1e**), consistent with its more uniform crystal growth [10, 32]. In contrast, RB@UiO-67 synthesised through the two-acid pathway exhibits more heterogeneity in crystal growth (**Fig. 1f**), indicating a defect-rich or kinetically trapped structure formed under non-equilibrium conditions where competing interactions from the guest may disrupt regular framework assembly [18]. Additional micrographs can be found in the Supporting Information (SI).

Sonochemically produced UiO-66 shows a near-normal distribution in particle size, consistent with the rapid and homogeneous nucleation induced by ultrasonic cavitation, followed by limited growth caused by the brief duration (**Fig. 1g**). By contrast, RB@UiO-67 nanocrystals exhibit a broader and more skewed particle size distribution spanning from tens to several hundreds of nanometres (**Fig. 1h**), indicative of competing nucleation and growth processes. The presence of luminescent guests during sonochemical synthesis of LG@MOF might perturb the coordination equilibrium and crystal growth pathway, leading to higher heterogeneity in crystal growth of the guest-encapsulated framework. Moreover, it is envisaged that ultrasonic cavitation could promote the formation of more defect-rich domains in UiO crystals compared with their solvothermal counterparts.

### 3.1.1 Effect of acid modulation on defect formation in UiO frameworks

Acid modulators influence the formation of UiO MOF structures through two related mechanisms: promoting linker deprotonation and controlling the crystallisation environment. The former depends primarily on acid strength (pKa, the pKa of HCl is around -7 to -8, and the pKa of $CH_3COOH$ is 4.76), which affects linker coordination to the $Zr_6$ clusters [33], whereas the latter is dependent on acid concentration, which regulates nucleation, crystal growth, and structural reorganisation. This distinction is consistent with coordination-modulation studies of Zr-based MOFs, in which modulator identity and concentration can influence defect formation through related but non-identical pathways [34].

In the absence of an added acid modulator, UiOs can still form due to the limited hydrolysis of DMF that provides trace $H^+$ species [32, 35], but with smaller crystallite sizes and less well-defined morphologies, indicating that insufficient acidity favours rapid nucleation rather than controlled crystal growth [18, 32]. This observation suggests that when the acidic component is insufficient, the reaction preferentially proceeds through linker deprotonation rather than effective control of crystallisation.

Based on this observation, a two-acid UiO synthesis protocol was developed to balance structural uniformity (See SEM images in SI, Section 1.2) and guest-loading capacity. In this approach, a weak acid was added in larger quantity to promote linker deprotonation, while a smaller amount of strong acid fine-tunes the crystallisation environment. The resulting UiO crystals exhibit improved faceting, larger crystallite sizes, and narrower diffraction peaks while requiring only approximately 25% of the reaction time needed to achieve comparable morphology using acetic acid alone. Enhanced structural control also improves guest loading and optical performance; under otherwise identical conditions, UiO-67 synthesised *via* the two-acid route confines more rhodamine B (RB) guests, producing a quantum yield 7.7% higher than that of the HCl-only system.

### 3.1.2 Comparison of UiO-66 with UiO-67

Powder X-ray diffraction patterns of UiO-66 together with the guest-free UiO-67 structural analyses (**Fig. 2a–g**) reveal that the two UiO frameworks respond differently to defect engineering despite sharing the same *fcu* topology. Rietveld refinement was performed on the PXRD patterns (**Fig. 2c**) to determine the lattice parameters and crystallographic densities used for defect analysis. Besides the crystallinity reflected by peak sharpness [36], weak low-angle reflections appearing after prolonged sonication suggest the formation of defect-ordered domains or supercell structures that have previously been reported in UiO frameworks [16, 32]. This trend is consistent with a shift in the Zr-terephthalate crystallisation pathway from the kinetically favoured UiO phase towards a more defective structure under continued ultrasonic reaction [37].

For UiO-66, the intensity and full width at half maximum (FWHM) ratios between the (111) and (200) reflections are resembling the solvothermal control sample (**Fig. 2d**). The small variation in intensity ratio indicates that ultrasonic synthesis has little influence on preferred crystallographic orientation, while the FWHM ratio remaining near unity suggests that defect formation is largely isotropic. Together, these observations indicate that acid modulation primarily changes the concentration of structural defects without introducing significant anisotropic microstructural distortion.

The lattice parameters and crystallographic densities obtained from Rietveld refinement further distinguish the two frameworks (**Fig. 2f, g**). Most sonochemically synthesised samples exhibit higher apparent densities than the ideal defect-free models, reflecting deviations from the ideal structure including reduced particle size and denser packing [38, 39]. In both UiO-66 and UiO-67, a negative correlation between lattice parameter and density is observed, consistent with the established mechanism that missing-linker or missing-cluster defects expand the lattice while reducing framework density. However, the dependence is more pronounced for UiO-67, indicating that the longer BPDC linker undergoes greater structural perturbation during defect formation than the relatively shorter BDC linker in UiO-66.

The role of acid modulators is further reflected in the progressive increase in defect concentration with increasing acid content. This behaviour is consistent with coordination modulation, in which modulators compete with dicarboxylate linkers for coordination to $Zr_6$ clusters and promote missing-linker defects. In contrast, synthesis without acid produces denser materials with smaller lattice parameters and higher framework connectivity. Extended sonication further enhances defect formation, likely through cavitation-induced uneven crystal growth and repeated dissolution-recrystallisation processes, while the two-acid protocol produces higher defect concentrations through the combined effects of enhanced deprotonation and multiple competing coordination species [40].

The influence of defect engineering on accessible porosity is reflected in the nitrogen adsorption isotherms and BET surface areas (**Fig. 2h**). For the guest-free materials, acid modulation substantially increases nitrogen uptake compared with sample obtained from the acid-free synthesis (designated as "no acid" in **Figs.2** and **3**), demonstrating that controlled defect formation enhances accessible pore volume without destroying framework integrity. The HCl-modulated UiO-66 samples exhibit BET surface areas of 944–1167 $m^2\ g^{-1}$ compared with 485 $m^2\ g^{-1}$ for the "no acid" sample, while the 2-acid protocol achieves 1079 $m^2\ g^{-1}$. Likewise, defect engineering through linker deficiency successfully increases porosity, as evidenced from the higher surface area of the "0.5L" sample synthesised with an M:L ratio of 1:0.5 (1167 $m^2\ g^{-1}$) compared with the excessive-linker sample denoted as "1.3L" (944 $m^2\ g^{-1}$). These observations agree with previous studies showing that controlled vacancy formation enhances accessible pore volume in UiO-type frameworks [19, 41].

A marked reduction in nitrogen uptake is observed after encapsulation of rhodamine B and fluorescein (**Fig. 2i**). The decrease is substantially greater than the variation introduced by acid modulation alone, indicating that the luminescent guests occupy the internal microporous volume rather than merely adsorbing onto external particle surfaces.

The ATR-FTIR spectra provide complementary evidence for the local structural changes associated with defect formation (**Fig. 2j**). Variations in the carboxylate stretching region (~1400–1650 $cm^{-1}$) indicate changes in linker coordination and terminal hydroxyl environments, where broader bands correspond to a wider distribution of bonding states arising from missing-linker defects and hydrogen bonding, whereas sharper bands indicate more uniformly coordinated linkers [42, 43]. Correlating these spectroscopic trends with the X-ray diffraction results and FESEM imaging (**Fig. S6** in SI) suggests that the 2-acid protocol produces a uniform microstructure while preserving defect sites favourable for guest encapsulation.

Taken together, the diffraction data and Rietveld refinement, porosity and spectroscopic analyses demonstrate that controlled defect engineering benefits both UiO frameworks. Henceforth, the greater lattice flexibility and larger pore architecture of UiO-67 could enable a higher guest encapsulation while maintaining structural integrity, supporting its selection as the host framework for luminescent guest confinement. (See SI, Section 3.1 for comparison between LG@UiO-66 and LG@UiO-67.)

### 3.1.3 Defect engineering for LG@UiO-67

A different behaviour was observed for LG@UiO-67. Solvothermal control results in severe peak merging and a small peak at lower angles (**Fig. 3a**), indicating the formation of supercells over extended heating [16, 44], while the 2-acid protocol sample had intensity and FWHM ratios resembling the simulated structure. Higher acid loadings also led to partial merging of the (111) and (200) reflections, arising from significant peak broadening, which is commonly associated with reduced crystallite size, increased defect concentration, and guest-induced structural disorder [45, 46].

Powder diffraction patterns (**Fig. 3a**) demonstrate that LG@UiO-67 fabricated sonochemically exhibits better crystallinity than the solvothermally synthesised congener, evidenced by the sharper separation of the (111) and (200) reflections. In contrast, the solvothermal sample shows severe peak merging together with a weak low-angle diffraction peak, suggesting the formation of defect-ordered supercell domains during prolonged heating. Among the sonochemical samples, the 2-acid protocol produces diffraction features most closely resembling the simulated UiO-67 structure, indicating that controlled acid modulation improves long-range structural order while maintaining defect formation.

To evaluate the microstructural evolution, the intensity and FWHM ratios of the (111) and (200) reflections were analysed (**Fig. 3b**). The solvothermal sample exhibits the largest intensity ratio, indicating a stronger preferred orientation towards the (111) lattice, whereas all sonochemical samples remain substantially closer to unity. The corresponding FWHM ratios likewise tend to unity for the 2-acid sample, suggesting reduced orientation-dependent peak broadening. Higher HCl content in synthesis broadens both reflections, consistent with reduced crystallite size, increased defect concentration, and guest-induced structural disorder.

Among the sonochemical groups, the 2 mL HCl and 2-acid samples exhibit the largest lattice parameters (**Fig. 3c**) without significantly compromising crystallinity. The expanded unit cells are consistent with increased missing-linker defect formation while retaining framework integrity, indicating that acid modulation promotes defect generation alongside controlled crystal growth.

The defect chemistry suggested by PXRD is further supported by ATR-FTIR spectroscopy (**Fig. 3d**). Systematic variations are observed in the carboxylate stretching region (~1400–1650 $cm^{-1}$), where sharper and better-defined bands correspond to more uniformly coordinated BPDC linkers, while broader features indicate a wider distribution of bonding environments associated with missing-linker defects and hydrogen-bonded terminal hydroxyl or water species. Consistent with the diffraction results, the 2-acid sample exhibits a similar relative peak intensity. The data suggest that the combined modulator strategy improves the underlying framework uniformity while preserving defect sites beneficial for guest encapsulation study.

Thermogravimetric analysis (TGA) further supports these structural differences (See the TGA curves of UiO-66 in SI, Section 2.3.). **Figs. 3e, f** show two major mass-loss events centred at approximately 210 °C and 570 °C, corresponding respectively to the decomposition of the encapsulated rhodamine B/fluorescein guests and the UiO-67 framework linker. The relative intensity of the first derivative peak indicates that the no-acid sample possesses the highest linker-to-guest ratio, suggesting the lowest guest-loading capacity despite its high defect content. This agrees with its comparatively small lattice parameter obtained from Rietveld refinement. Conversely, although the solvothermal sample displays the lowest linker-to-guest ratio, its merged diffraction peaks and relatively small lattice parameter (**Fig. 3c**) indicate that neither crystallinity nor pore development is optimised. The guest loading was also estimated from the TGA curves [47]. We found approximately 6.4 wt.% for RB and F confined in UiO-66, and 10.5 wt.% for RB and F encapsulated inside UiO-67 (see SI, Section 2.3). In contrast, a physical mixture of the matching concentration of the guests and a non-luminescent "host" (e.g. $BaSO_4$) only led to a non-emissive material due to AIQ.

Overall, the 2-acid protocol provides the best combination of structure and properties, and subsequently the nanoconfinement of luminescent guests.

### 3.2 Luminescence studies of UiO MOFs and LG@UiOs

UiO-67 was selected for optical properties characterisation owing to the stronger luminescence of the BPDC linker relative to UiO-66. Excitation and emission spectra of the UiO host synthesised under different conditions were compared (**Fig. 4a**), revealing that longer reaction time lowers the QY of UiO-67, because of the quenching of linker luminescence in closer packing [48]. Also, the 2-acid protocol significantly caused a red-shift in the emission of UiO-67. However, this phenomenon is reversed in the presence of a red-emitting guest (RB) (**Fig. 4b**), where the 2-acid protocol exhibits a higher QY suggesting improved guest encapsulation. Besides, sonochemically synthesised RB@UiO-67 exhibits a blue shift in emission but not excitation from its solvothermal counterpart, as longer ultrasonication time leads to red-shifted emission, it is hypothesised that prior to the degradation of dyes under high temperatures, longer reaction time results in higher guest loading. Also, although higher RB load promotes the red emission in the composite, it also leads to aggregation-induced quenching (AIQ) and compromises the emission intensity. The effect of acid amount on photophysical properties has been studied with F@UiO-67 (**Fig. 4c**). As the UiO-67 host emits at shorter wavelengths than fluorescein (F) does, the red shift observed in the 4 mL HCl group verified the hypothesis that a more defective MOF can confine more guests and thus, achieve stronger emission associated with the guest.

The excitation–emission maps revealed broad emission bands whose profiles vary systematically with synthesis conditions (**Fig. 4d, e**), indicating a strong influence of defect structure on the emissive behaviour. Consistently, fluorescence lifetimes and quantum yields vary across the UiO-67 series (**Fig. 4f**). Despite relatively similar lifetimes, substantial differences in quantum yield (QY) are observed, suggesting that defect formation primarily alters non-radiative decay pathways [28]. Samples synthesised using shorter reaction times and the 2-acid protocol exhibit the highest quantum yields, indicating that controlled defect formation can enhance emission efficiency [16, 19, 35], which is also proven with the improved quantum yields of UiO-67 in comparison to its linker BPDC.

The corresponding emission colours are visualised in the CIE 1976 diagram (**Fig. 4g**), where different synthesis conditions occupy distinct regions of the colour space. The shift in emission coordinates relative to the BPDC linker demonstrates that the luminescence of Zr MOFs integrating fluorescent linkers is not solely determined by the organic linker but is also strongly affected by the framework environment and defect structure. Also, introducing a blue-emitting guest into the blue-emitting UiO-67 host did not improve quantum yield, likely due to spectral overlap between guest and host absorption and emission bands. UiO-67 was therefore employed as the blue–purple colour emitter in the prototype SSL device, see section 3.3 below.

Subsequently, the RB and F guests were selected to demonstrate the feasibility of covering the whole visible spectrum by confining mixed guests in the UiO-67 host, accomplishing different mixing ratios (Fig. 4h).

Results of the accelerated photostability test demonstrate excellent long-term photostability against ultraviolet (UV) irradiation (**Fig. 4i, j**). RB+F@UiO-67 retained 81% of its initial quantum yield after 15 months of storage in ambient air, it maintained 70% of its initial emission intensity after 7200 min of continuous excitation at 560 nm and exhibited only a 3% decline in quantum yield after 1080 min of operation under the SSL device excitation source (**Fig. 4j**). Both the emission intensity and quantum yield data show a plateauing trend, indicating the presence of a photostable emissive species shielded by the UiO-67 host. Such behaviour is consistent with heterogeneous confinement environments within the LG@MOF composites, where less protected dye molecules will decay first, while those more effectively confined within the pores are shielded from UV degradation for extended irradiation.

### 3.3 Proof-of-concept SSL device incorporating LG@MOF

The ‘rainbow’ SSL demonstration slide (**Fig. 5a**) was made by depositing UiO-67 and LG@UiO-67 samples emitting in seven different colours, serving as a visual proof-of-concept of what the SSL device can achieve in terms of tuneable emission colours. The colours of the rainbow, both under ambient light and under UV, are tuned by adjusting the type and guest loading of fluorophores encapsulated in the MOF host, demonstrating that solid LG@MOF powders can act as wavelength-converting filters for UV excitation with controllable colour output and intensity. **Fig. 5b** shows a schematic of the SSL device, where individually prepared LG@UiO-67 samples were physically mounted as interchangeable thin-film filters on top of a UV 395 nm diode, enabling flexible tuning of the emitted light employing a constant emission source.

The optical performance of the ‘rainbow’ samples is further reflected in the quantum yield values (**Fig. 5c**), where the variation across the series corresponds to diversity in guest type and loading. The emission profiles (**Fig. 5d**) confirm that the colours observed arise from distinct and tuneable emission bands, with their positions shown in the CIE 1931 chromaticity diagram (**Fig. 5e**). Together, the preliminary results revealed the potential for tuning the photophysical properties and controlling the SSL device performance employing the LG@UiO-67 composites. Moreover, emission in the continuous spectral range between red and green can be achieved by physically layering thin films of LG@UiO-67, as exemplified in **Figs. S20 and S21**.

## 4 Conclusions

This study has demonstrated the facile sonochemical synthesis of luminescent Zr-based MOFs (UiO-66 and UiO-67) and their defect-engineered LG@MOF composites through one-pot encapsulation of luminescent dyes. Ultrasonication has accelerated the fabrication of MOFs and LG@MOFs by more than 90% and helps to achieve smaller and more faceted nanocrystals (See SI, Section 1.1 for additional AFM images of sonochemically synthesised LG@UiO particles.) suitable for luminescent inks and solid-state light emitters.

A two-acid sonochemical UiO synthesis protocol has been developed. This strategy produces well-faceted UiO crystals with controlled defect formation, enhanced porosity, and improves guest-loading capacity with a substantially shortened reaction time compared with the conventional single-acid method. The comparative investigation of UiO-66 and UiO-67 further demonstrated that despite the same topology, UiO-67 offers greater lattice flexibility and higher porosity, enabling better suppression of quenching, and tuneable solid-state emission across the visible spectrum.

This study also provides insights into the interplay between synthetic factors, structural defects, and photophysical behaviours of LG@Zr-MOF materials. Defect engineering has been demonstrated to regulate guest loading, and thus emission colour and quantum yield. Furthermore, the enhanced long-term photostability of LG@UiO-67 demonstrates the protective effect of the UiO host and the scalability of sonochemical synthesis for bio-compatible luminescent MOF composites.

A proof-of-concept SSL device has been demonstrated, leveraging the tuneable emission and stability of the LG@Zr-MOF nanocomposites. Prospective applications include solid-state lighting, chemosensing, and bioimaging. Future studies will explore the effect of guest–host interactions, defect control for optical sensing, and accelerated sonochemical manufacturing of other families of MOF-based composites.

## Supporting Information (SI)

The SI contains additional AFM images, SEM micrographs, Scanning Transmission Electron Microscope (STEM) images, XRD patterns, Rietvield refinement analysis, ATR-FTIR spectra, TGA plots, excitation-emission spectra, quantum yield data, CIE plots, Energy-Dispersive X-ray Spectroscopy (EDS) plots, and more demonstration of the SSL device.

## Acknowledgements

J.-C.T. thanks the Engineering and Physical Sciences Research Council, EPSRC (EP/R511742/1) for equipment funding.

## Author Contributions

Conceptualisation and methodology were developed by Y.W. and J.-C.T. Materials synthesis, sample characterisation and subsequent data analyses were performed by Y.W. under the guidance of J.-C.T. Nitrogen adsorption and thermogravimetric measurements were performed by J.D. Taylor. Manuscript was drafted by Y.W., edited by J.-C.T., and subsequently reviewed by all authors. Project supervision and equipment funding acquisition by J.-C.T.

## Data Statement

The data that support the findings of this study are available in the supplementary material of this article. Any other data will be made available on request.

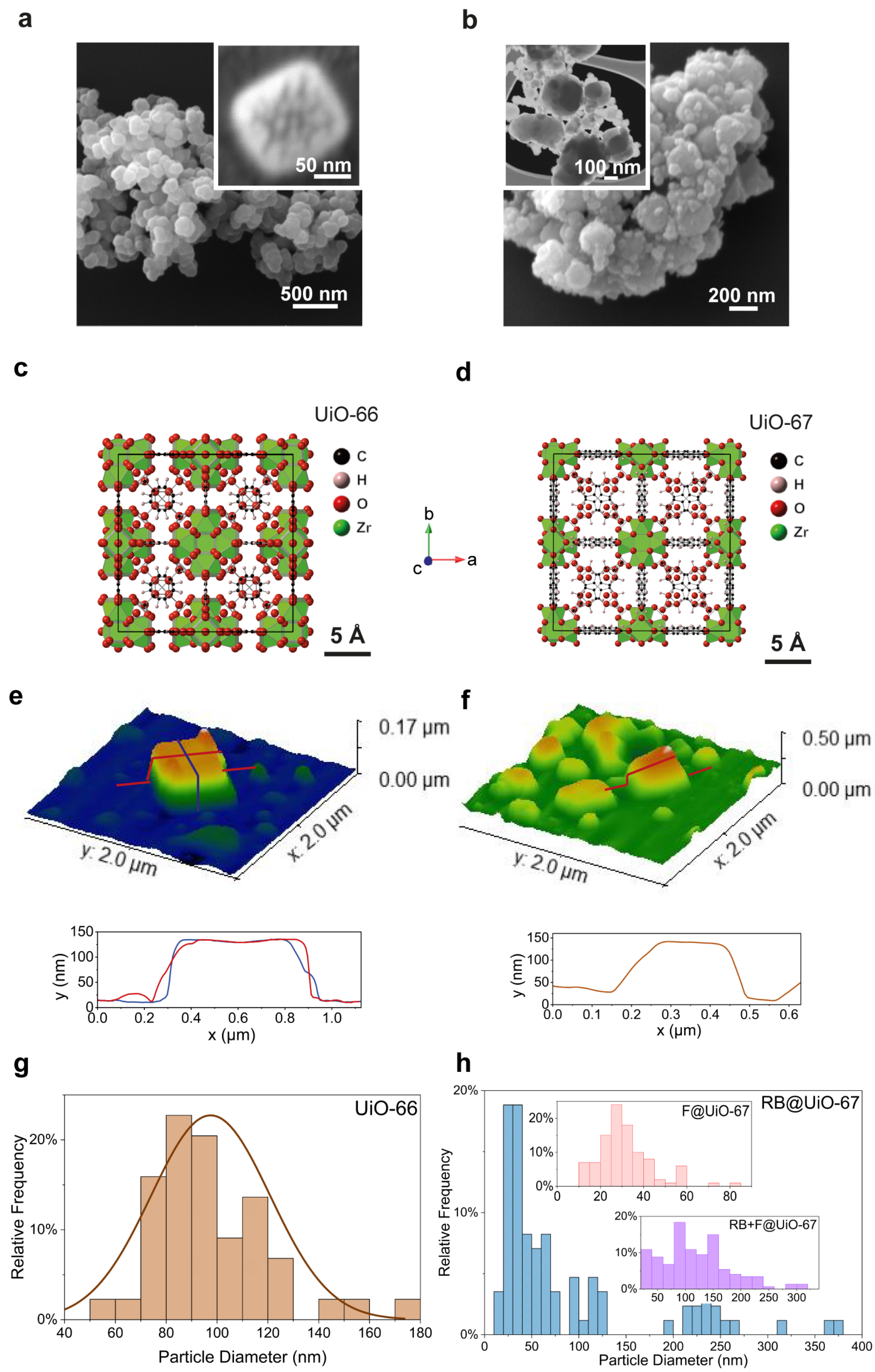

**Fig. 1 (a)** SEM micrographs of sonochemically-synthesised UiO-66 nanocrystals. **(b)** SEM and STEM images of sonochemically-synthesised RB@UiO-67 employing the two-acid protocol. **(c, d)** Crystal structure of a unit cell of UiO-66 and UiO-67, respectively. **(e, f)** AFM height topography and cross-sectional profiles of sonochemically-synthesised UiO-66 and RB@UiO-

67, respectively. **(g)** Particle size distribution of sonochemically-synthesised UiO-66 (sample size, *n* = 100). **(h)** Particle size distribution of sonochemically-synthesised RB@UiO-67 (*n* = 200), F@UiO-67 (*n* = 100), and RB+F@UiO-67 (*n* = 150).

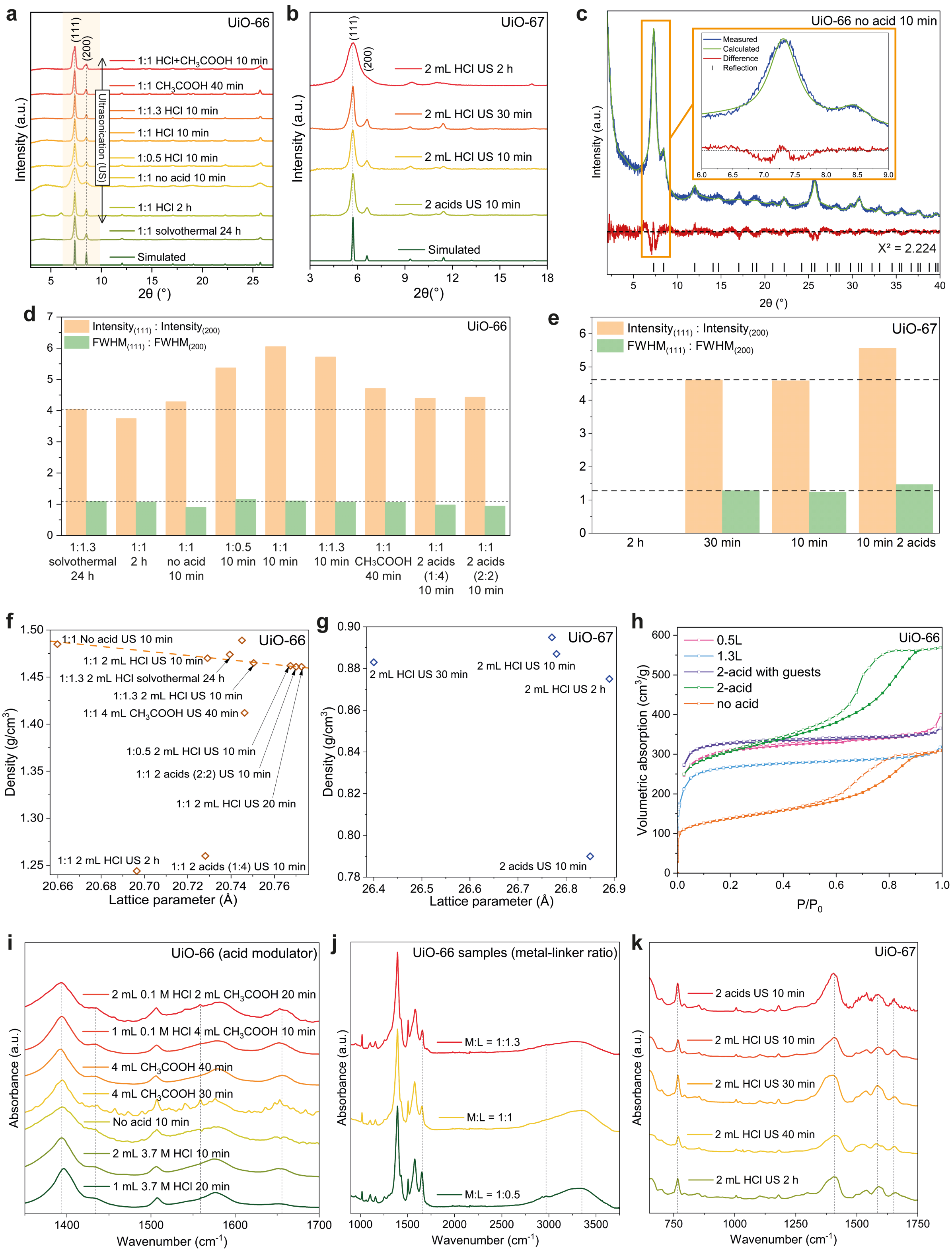

**Fig. 2** **(a, b)** Powder X-ray diffraction (PXRD) patterns of UiO-66 and UiO-67 samples. The simulated XRD patterns are calculated from CCDC 1018045 (UiO-66) and 1018032 (UiO-67). **(c)** An exemplar of Rietveld refinement with UiO-66. **(d, e)** Histogram of the intensity and full width at half maximum (FWHM) ratios between (111) and (200) peaks in UiO-66 and UiO-67

samples, respectively. Note that all patterns were baseline corrected prior to peak analysis. **(f, g)** Lattice parameters – density scatter plot of UiO-66 and UiO-67 samples calculated by Rietveld refinement. The dotted lines are guides for the eye.

*Note: the sample synthesised with M: L = 1: 1.3 is denoted as 1.3L, the sample synthesised with M: L = 1: 0.5 is denoted as 0.5L. See more demonstration on Rietveld refinement in SI Section 2.2.

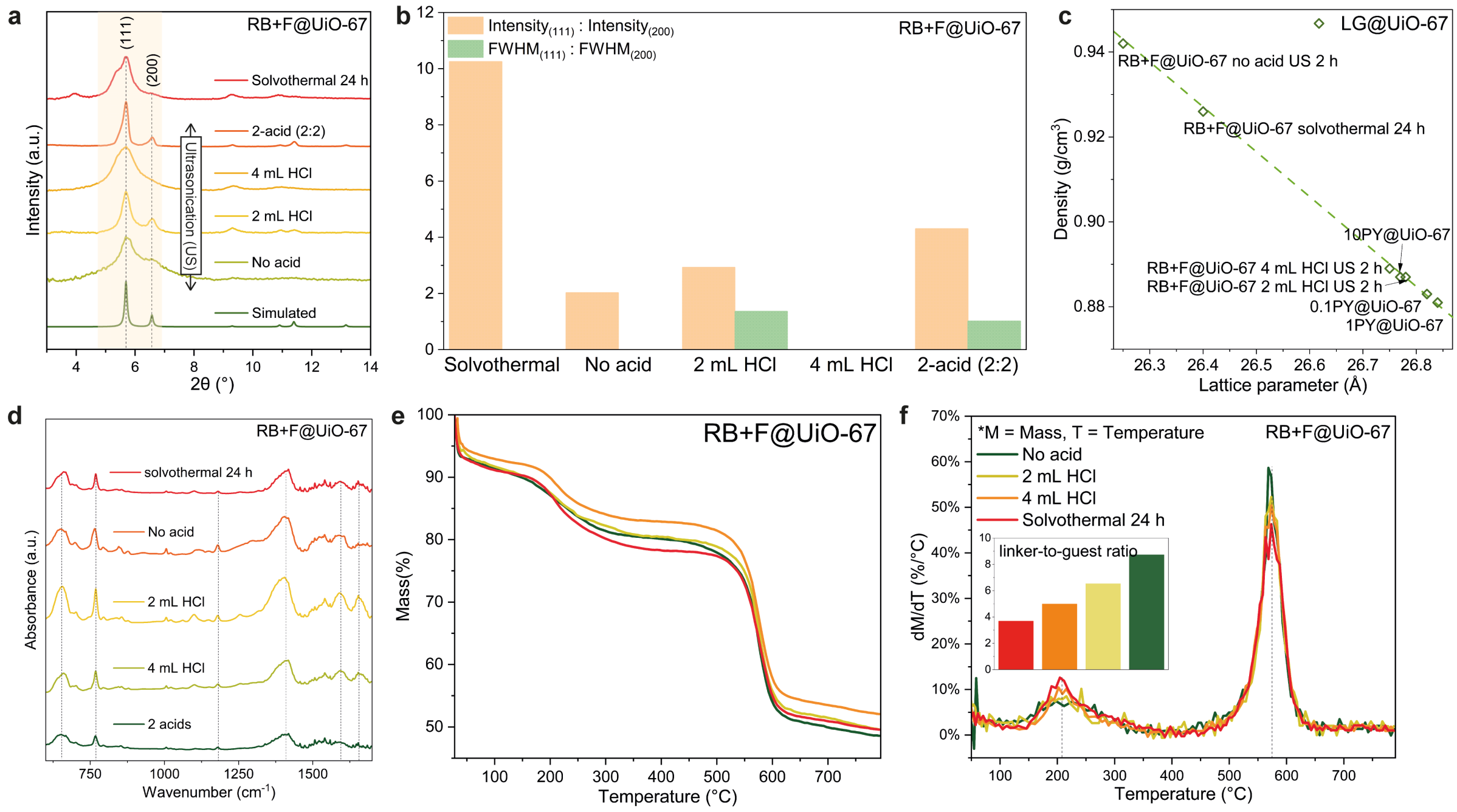

**Fig. 3 (a)** **P**XRD patterns of LG@UiO-67 samples made with different energy sources, ultrasonication time, and acid modulators. The simulated XRD pattern is calculated from CCDC 1018032. **(b)** Histogram of the intensity and full width at half maximum (FWHM) ratios between (111) and (200) peaks in RB+F@UiO-67 samples. **(c)** Lattice parameters – density scatter plot of UiO-67 and LG@UiO-67 samples calculated by Rietveld refinement. Note: PY is pyrene, a purple-emitting guest. **(d)** ATR-FTIR spectra of RB+F@UiO-67 samples made with different energy sources and acid modulator amounts. **(e)** TGA curves of RB+F@UiO-67 samples synthesised under different ultrasonication conditions. **(f)** Calculated from **(e)**, the rate of change in mass with temperature of the RB+F@UiO-67 samples.

*Note: For **Fig. 3c**, the calculated densities are for the host UiO-67 only but denoted with corresponding guests to demonstrate the effect of guest presence in the *in situ* synthesis.

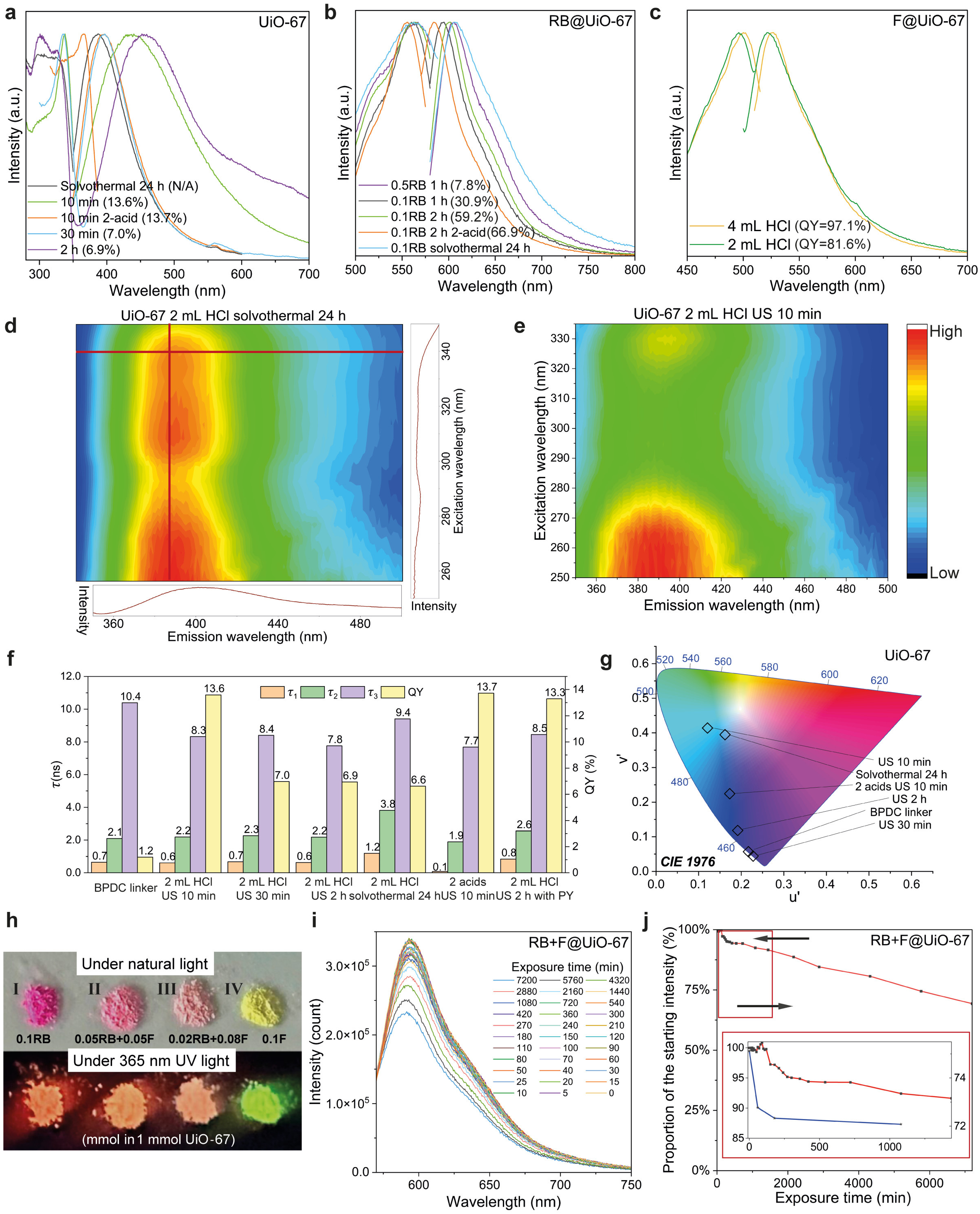

**Fig. 4 (a-c)** Excitation and emission profiles of UiO-67 and LG@UiO-67 samples with QY values indicated in the brackets. Note that QY 'N/A' indicates negative measured value that represents net absorption. **(d, e)** Excitation-emission maps of UiO-67 samples ultrasonically synthesised under different conditions. **(f)** Fluorescence lifetimes and quantum yield of BPDC linker and UiO-67 samples made with different reaction parameters. **(g)** CIE 1976 plot showing the colours of the emission of UiO-67 and its linker. **(h)** LG@UiO-67 powder samples under ambient light and excited under UV 365 nm. **(i)** Emission profiles of RB+F@UiO-67 measured during the accelerated photostability test. **(j)** Variation of emission intensity (red) and quantum yield (green) of RB+F@UiO-67 as a function of time.

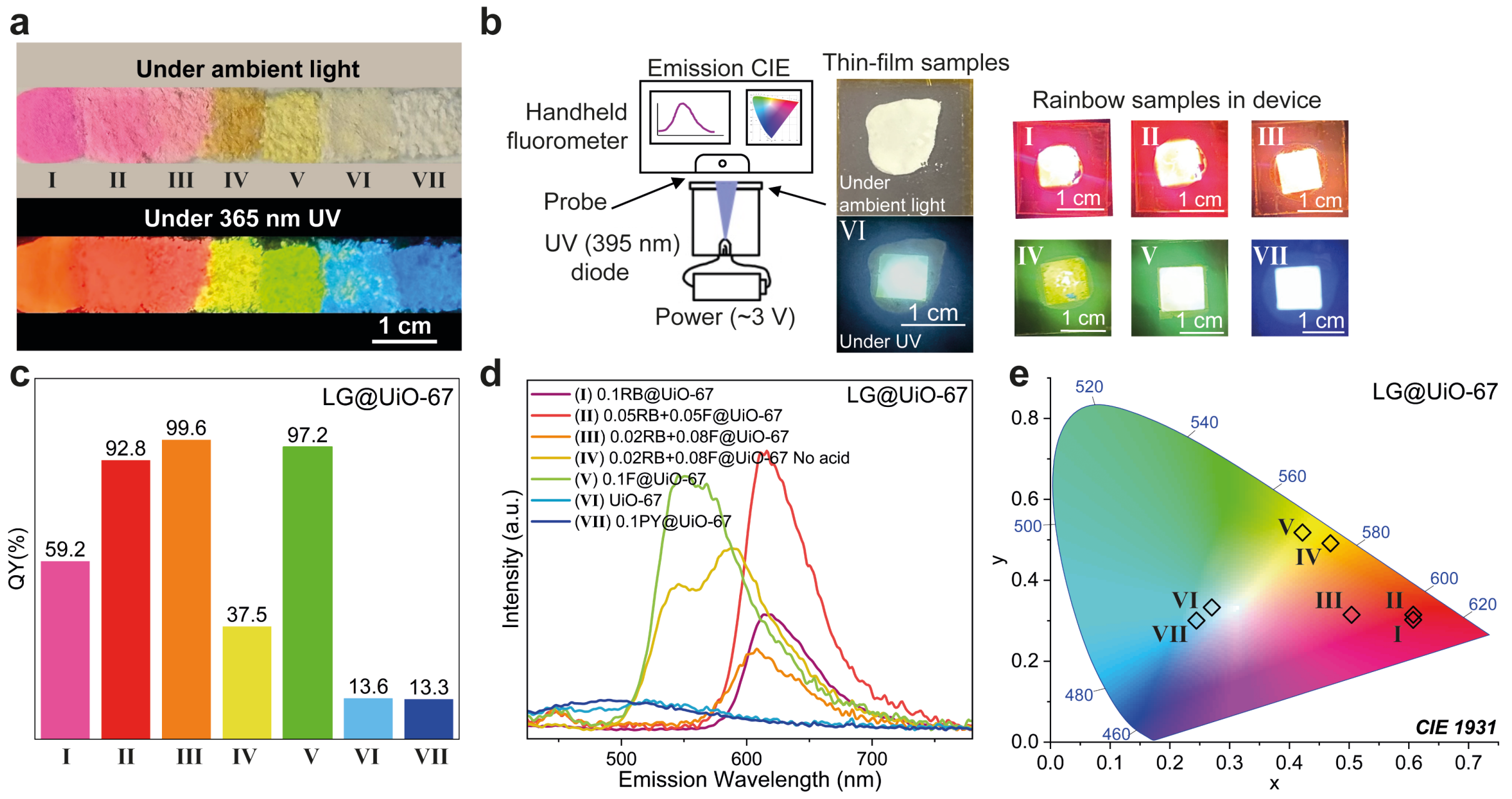

**Fig. 5 (a)** The ‘rainbow’ solid-state lighting (SSL) sample slide. **(b)** A schematic of the SSL device setup. **(c)** Quantum yield values of the samples in panel (a). **(d)** Emission profiles of the samples excited under UV 395 nm. **(e)** CIE 1931 plot of the ‘rainbow’ samples in panel **(a)**.

# *Supporting Information*

# *for*

# Ultrasound-Assisted Facile Nanoconfinement of Luminescent Guests in Zirconium Metal-Organic Frameworks

Yiyi Wang,[1] James D. Taylor,[2] and Jin-Chong Tan[1,*]

[1]Multifunctional Materials & Composites (MMC) Laboratory, Department of Engineering Science, University of Oxford, Parks Road, Oxford OX1 3PJ, U.K.
[2]ISIS Neutron and Muon Source, Science and Technology Facility Council, UKRI, Rutherford Appleton Laboratory, Chilton, Didcot OX11 0QX, United Kingdom.

[*]*Corresponding author: jin-chong.tan@eng.ox.ac.uk*

## §1 Microscopic characterisation of nanocrystals

### 1.1 Atomic force microscopy (AFM)

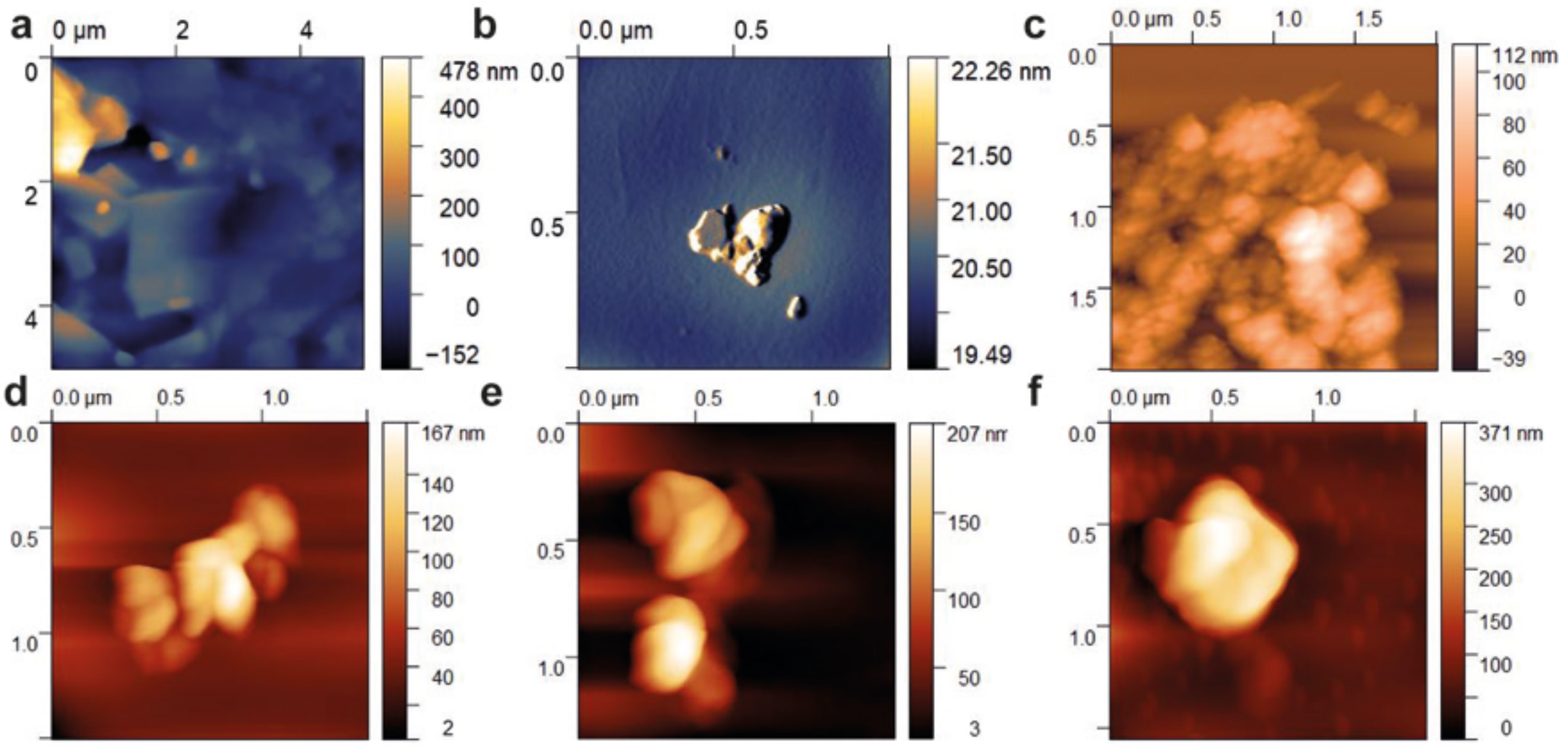

**Fig. S1** AFM height topography of **(a)** sonochemically (10 min) made UiO-66 crystals after 1 year of storage in ambient air. **(b)** UiO-66 sonochemically synthesised in 40 min with $CH_3COOH$ as the modulator. **(c)** UiO-66 sonochemically synthesised in 10 min that produced a gel product. **(d-f)** AFM images of 0.1PY@UiO-66 synthesised sonochemically in 2 h.

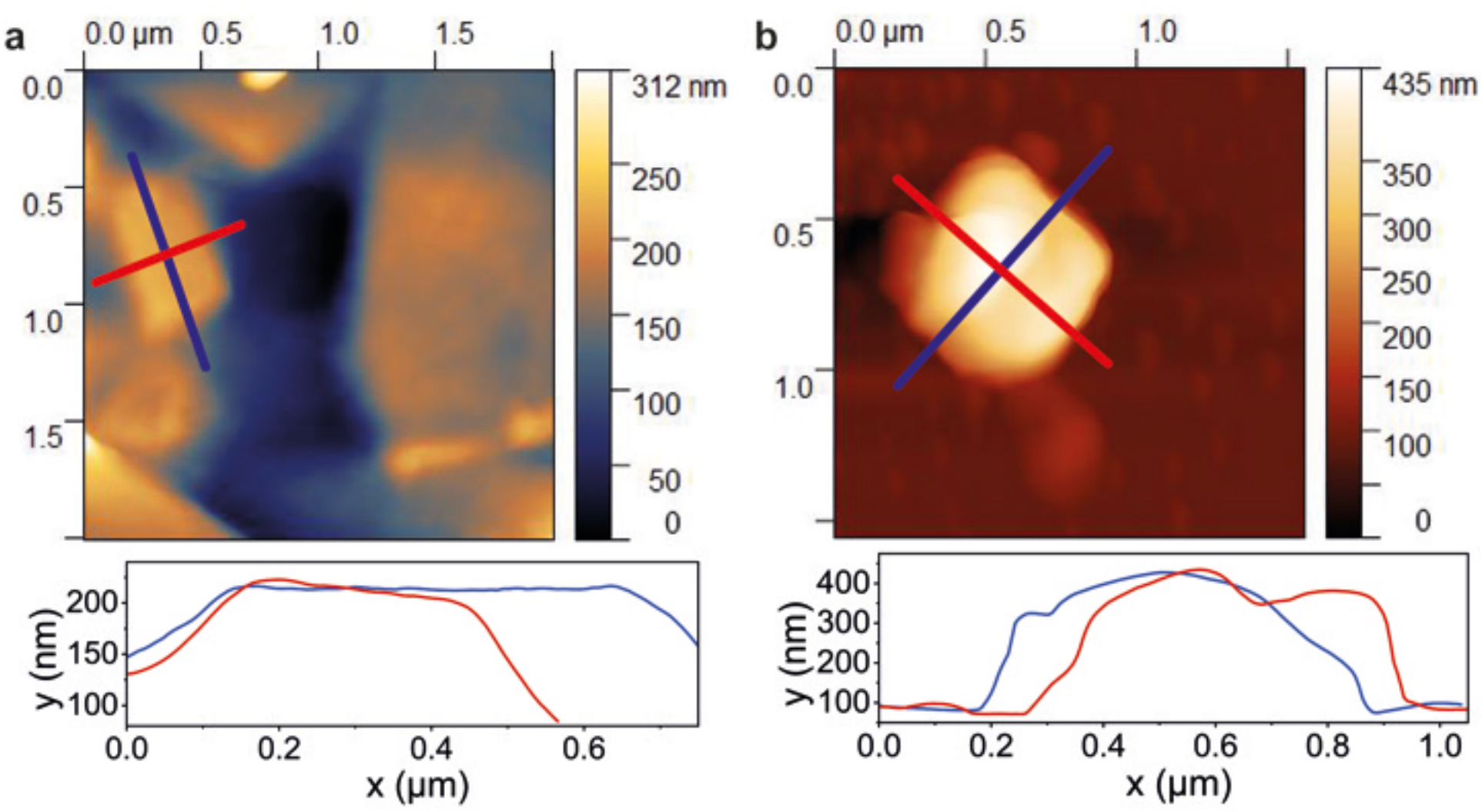

**Fig. S2** Cross-sectional plots of representative faceted particles in **(a)** UiO-66 sonochemically synthesised in 10 min with M: L = 1:1.3 and 2 mL HCl, stored for a year. **(b)** 0.1PY@UiO-66 synthesised sonochemically in 2 h.

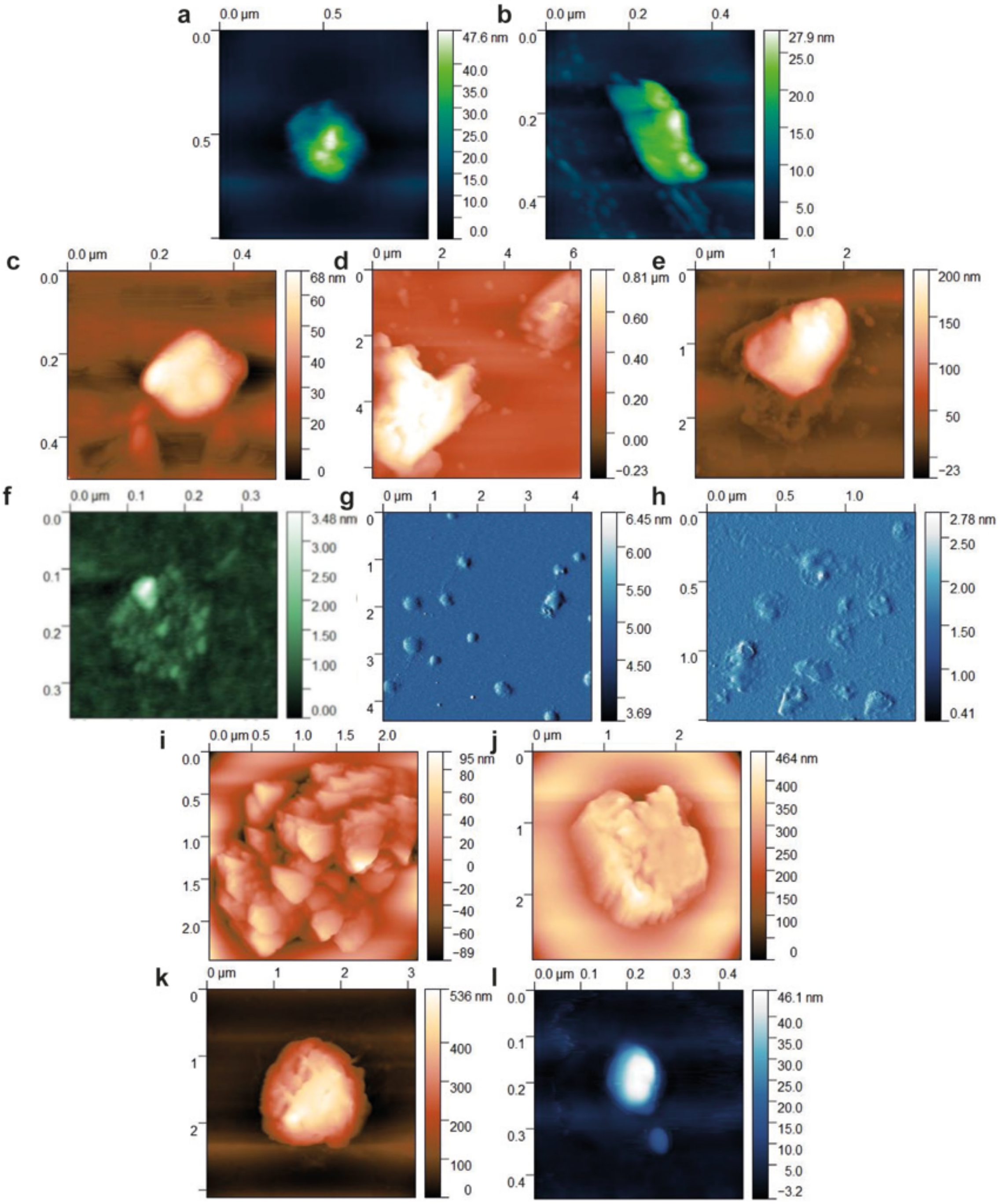

**Fig. S3** AFM images of **(a, b)** UiO-67 synthesised sonochemically in 10 min with the 2-acid protocol. **(c, d)** 0.1RB@UiO-67 (RB = rhodamine B, 0.1 = 0.1 mmol RB/mmol UiO-67 at mixing the reactants) synthesised sonochemically in 1 h. **(e)** 0.5RB@UiO-67 synthesised sonochemically in 1 h. **(f)** 0.1RB@UiO-67 synthesised sonochemically in 2 h. **(f)** 0.1F@UiO-67 synthesised sonochemically in 2 h. **(g, h)** 0.05RB+0.05F@UiO-67 synthesised sonochemically in 2 h. **(i)** 0.02RB+0.08F@UiO-67 synthesised with the 2-acid protocol. **(j)** 0.1RB@UiO-67 synthesised solvothermally in 24 h. **(k)** RB@UiO-67 synthesised with 30 min

of ultrasonication and 24 h of oven heating. **(l)** Perylene@UiO-67 (0.1PE@UiO-67) synthesised sonochemically in 2 h.

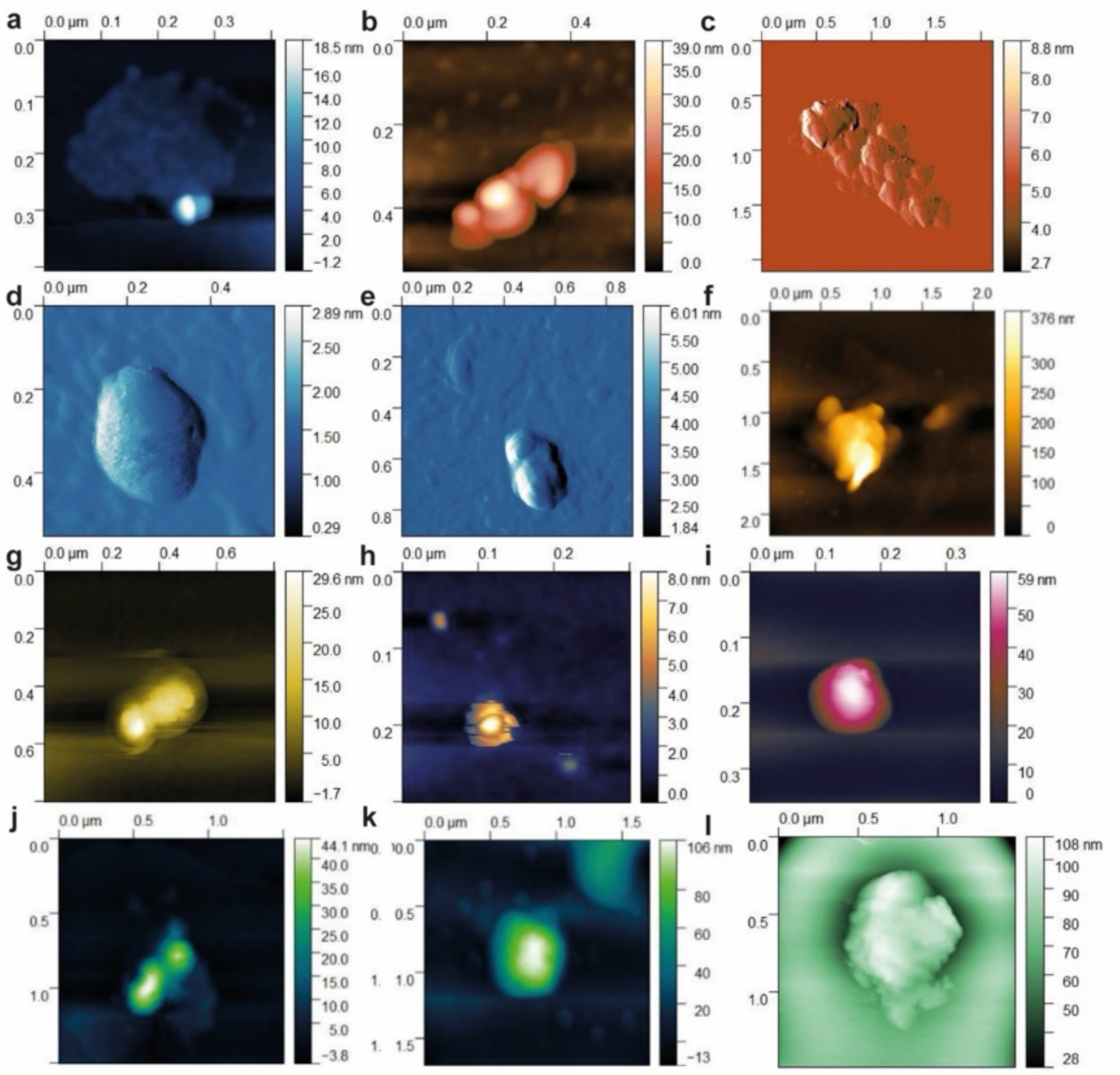

**Fig. S4 (a)** 0.1PE@UiO-6 synthesised sonochemically in 2 h. **(b, c)** 7-methoxycoumarin@UiO-67 (0.1MC@UiO-67) synthesised sonochemically in 2 h. **(d, e)** 7-0.1MC@UiO-67 synthesised sonochemically in 3 h. **(f)** 0.01MC@UiO-67 synthesised sonochemically in 2 h. **(g)** 0.001MC@UiO-67 synthesised sonochemically in 2 h. **(h, i)** Pyrene@UiO-67 (0.1PY@UiO-67) synthesised sonochemically in 2 h. **(j, k)** Methylene Blue@UiO-67 (0.1MB@UiO-67) synthesised sonochemically in 2 h. **(l)** AFM images of coumarin (C)@UiO-67 (0.1C@UiO-67) synthesised sonochemically in 2 h.

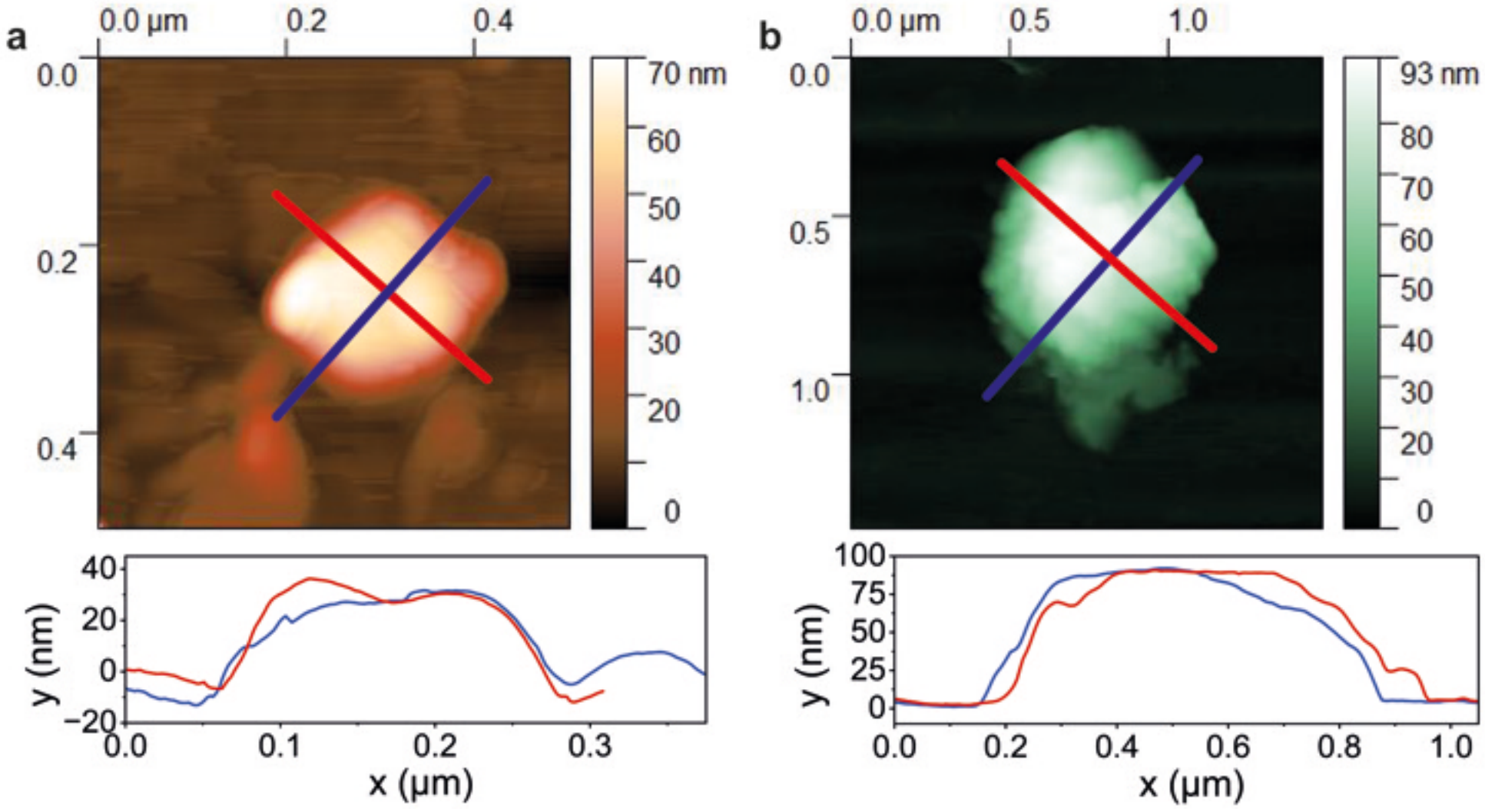

**Fig. S5** Cross-sectional plots of representative faceted particles in **(a)** 0.1RB@UiO-67 sonochemically synthesised in 1 h. **(b)** 0.1C@UiO-67 synthesised sonochemically in 2 h.

### 1.2 Scanning Electron Microscope (SEM) Images

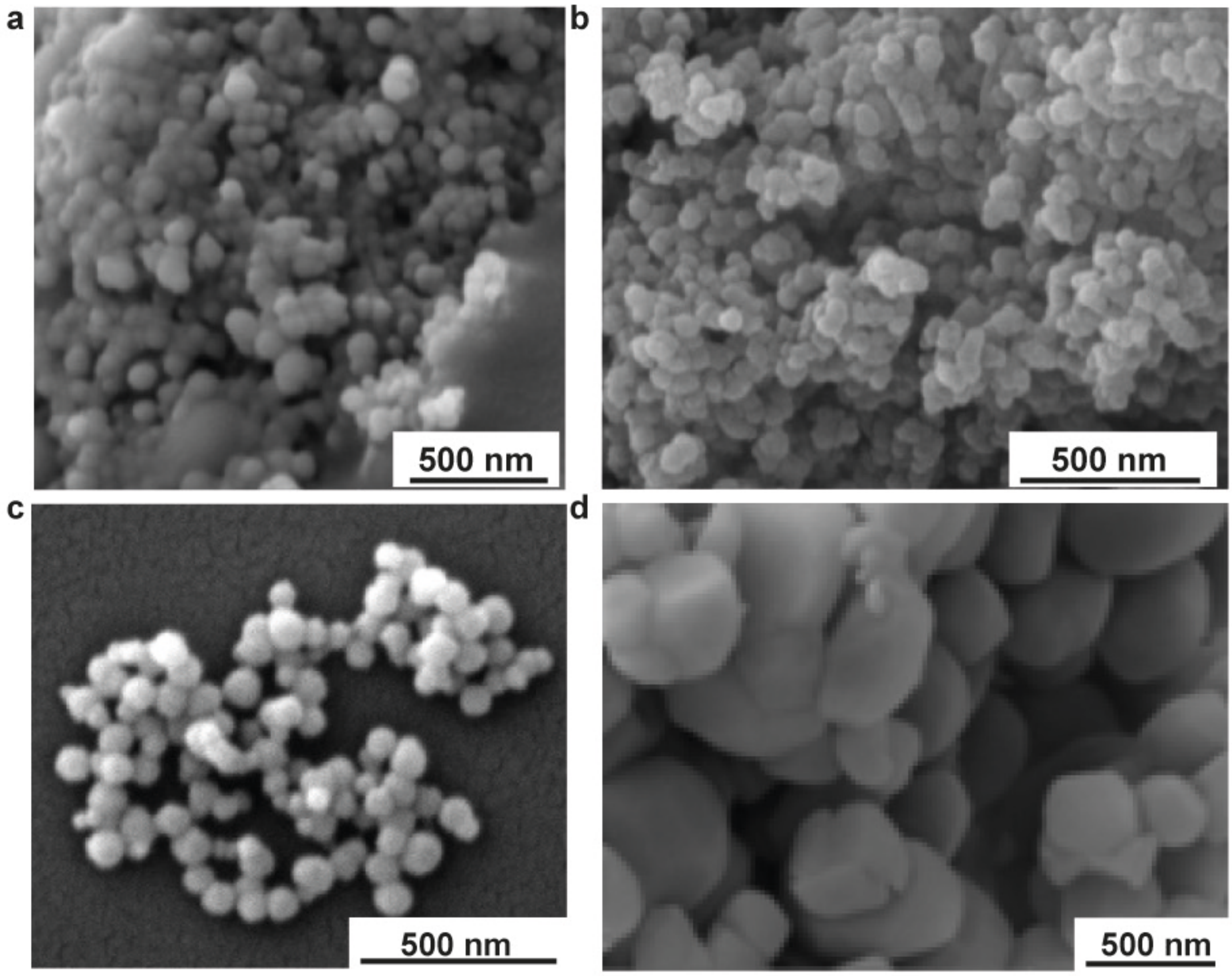

**Fig. S6** FESEM micrographs of **(a)** UiO-66 synthesised without acid modulation. **(b)** UiO-66 synthesised with acetic acid as the modulator. **(c)** UiO-66 synthesised with the 2-acid protocol. **(d)** Sonochemically synthesised 0.01RB+0.09F@UiO-66.

### 1.3 Scanning Transmission Electron Microscope (STEM)

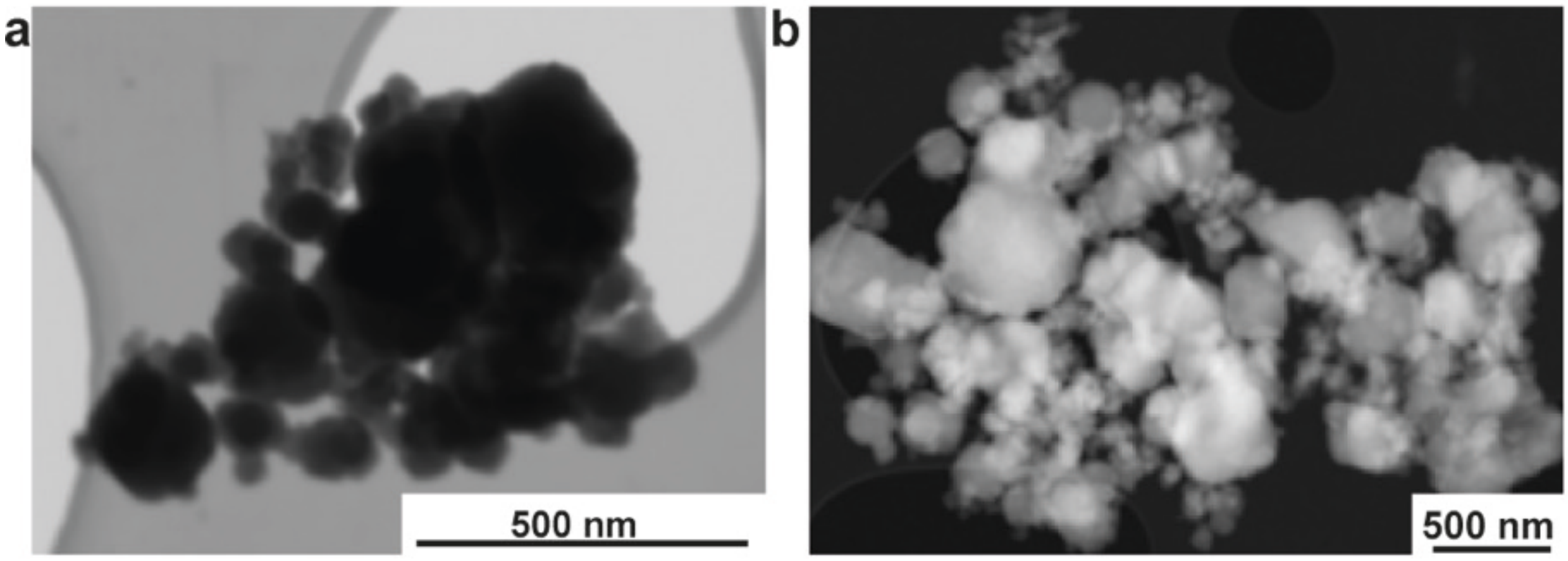

**Fig. S7** STEM images of **(a)** UiO-67 synthesised sonochemically in 10 min with the 2-acid protocol. **(b)** RB@UiO-67 synthesised sonochemically in 2 h with the 2-acid protocol.

## §2 Defect Analysis

### 2.1 X-Ray Diffraction (XRD) and Rietvield Refinement Demonstrations

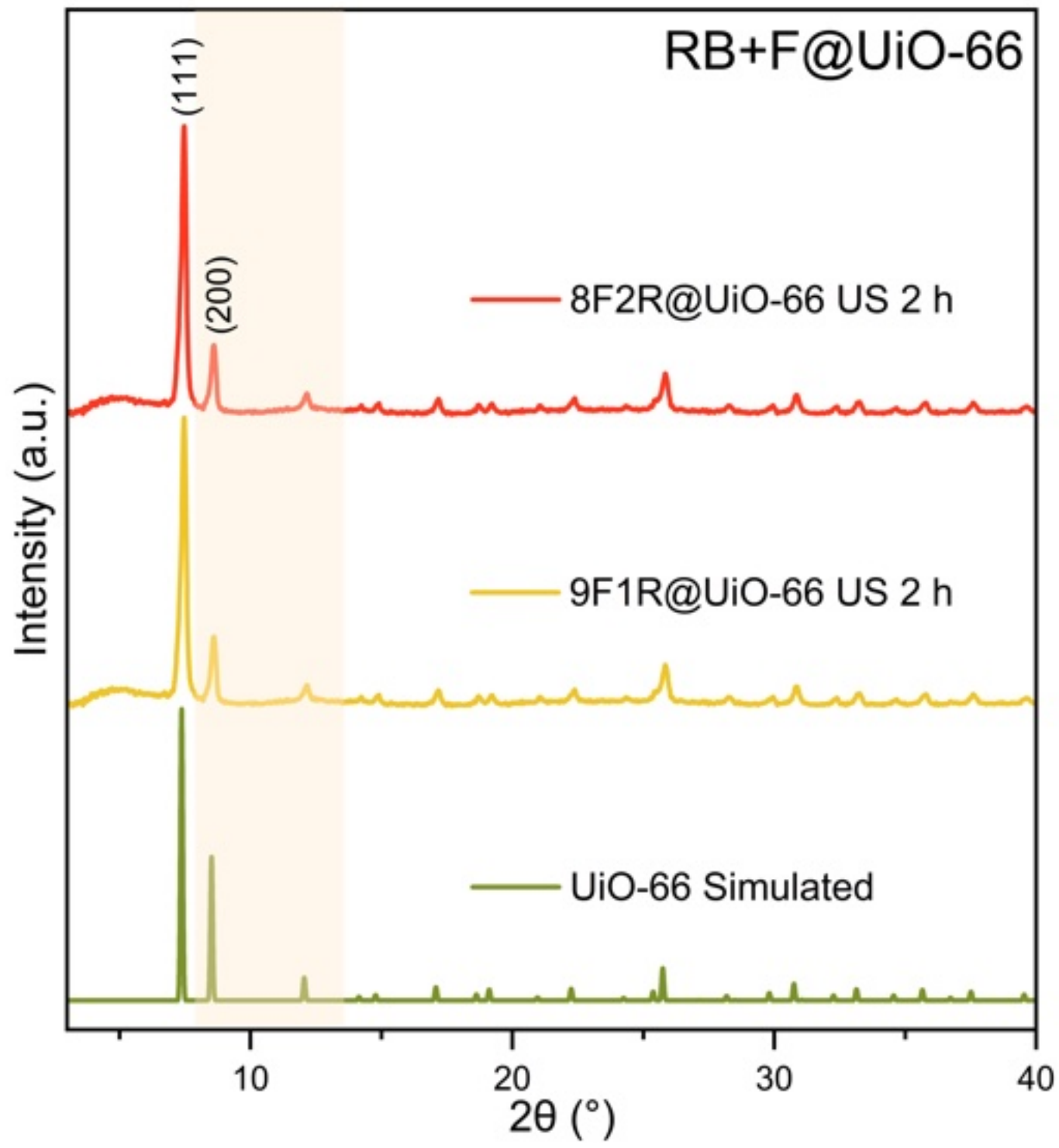

**Fig. S9** XRD patterns of sonochemically synthesised RB+F@UiO-66 samples compared with the XRD pattern of UiO-66 simulated from CIF [CCDC 1018045].

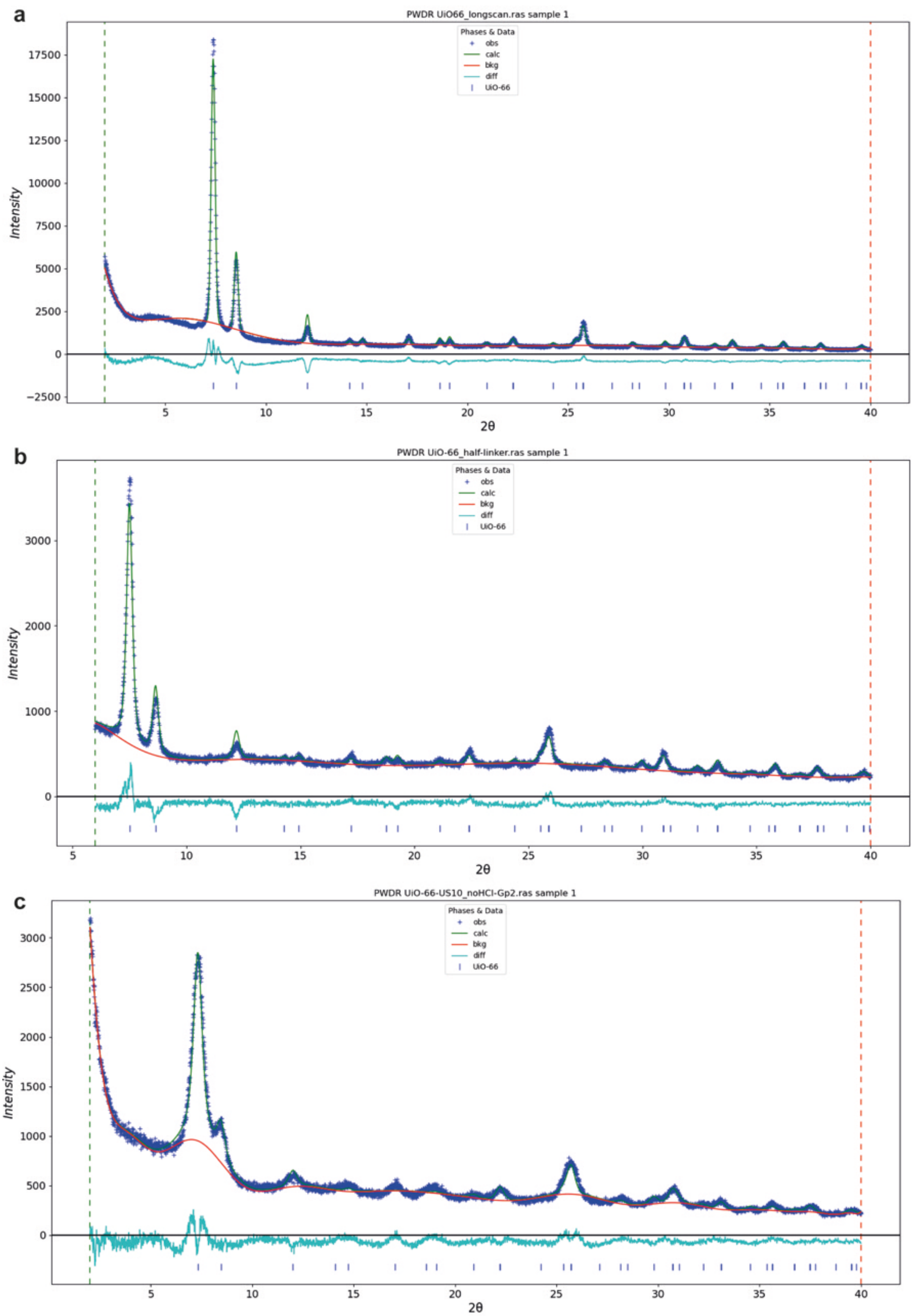

**Fig. S10** Demonstration of Rietveld refinements of the XRD patterns of **(a)** solvothermally synthesised UiO-66 ($X^2$ = 14.9). **(b)** UiO-66 synthesised with 10-minute ultrasonication and metal-linker ratio 1:0.5 ($X^2 = 2.3$). **(c)** UiO-66 synthesised with 10-minute ultrasonication and no acid modulation ($X^2 = 2.2$).

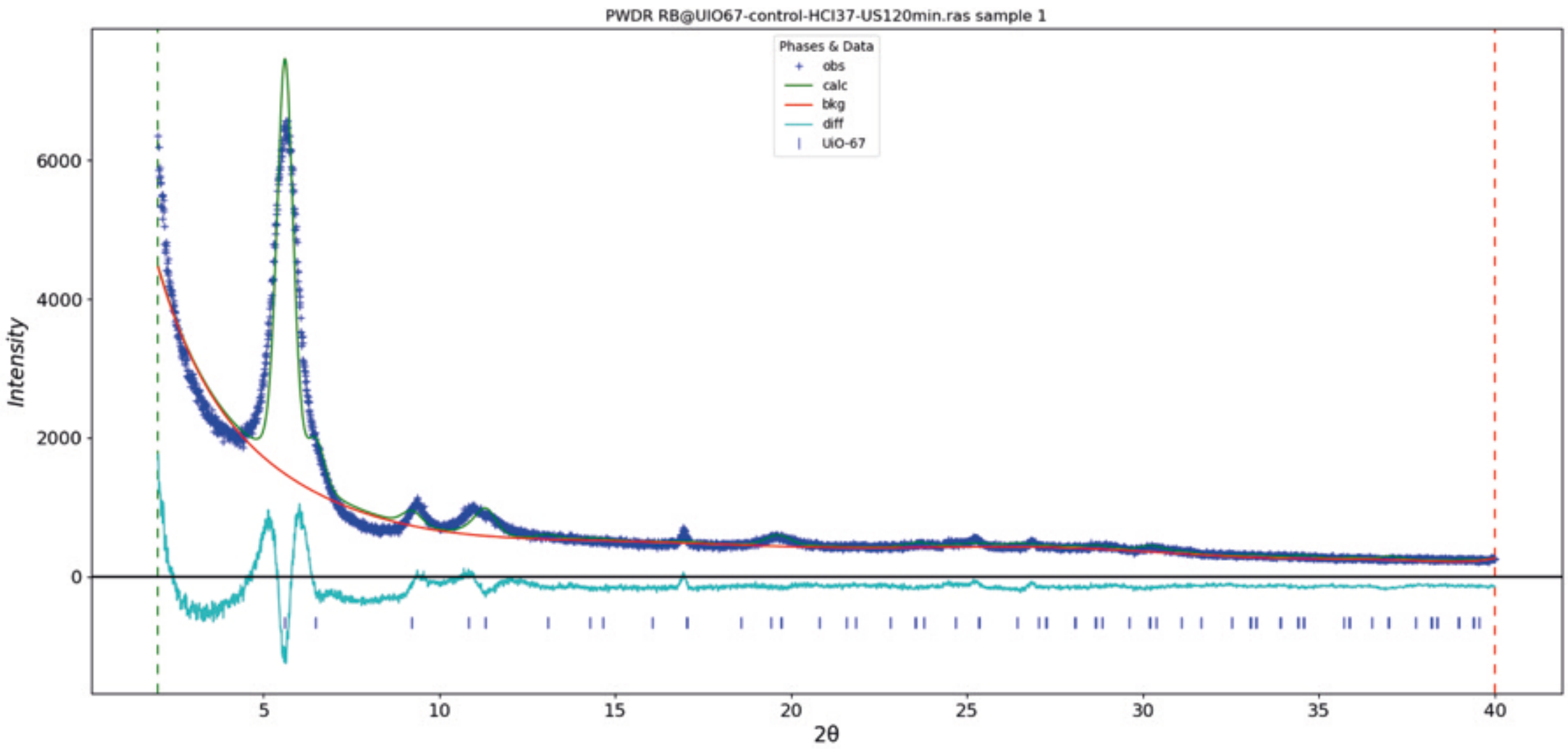

**Fig. S11** Demonstration of Rietveld refinements of the XRD patterns of UiO-67 synthesised sonochemically in 2 h ($X^2 = 14.1$).

## 2.2 Attenuated Total Reflectance Fourier-Transform Infrared (ATR-FTIR) Spectra

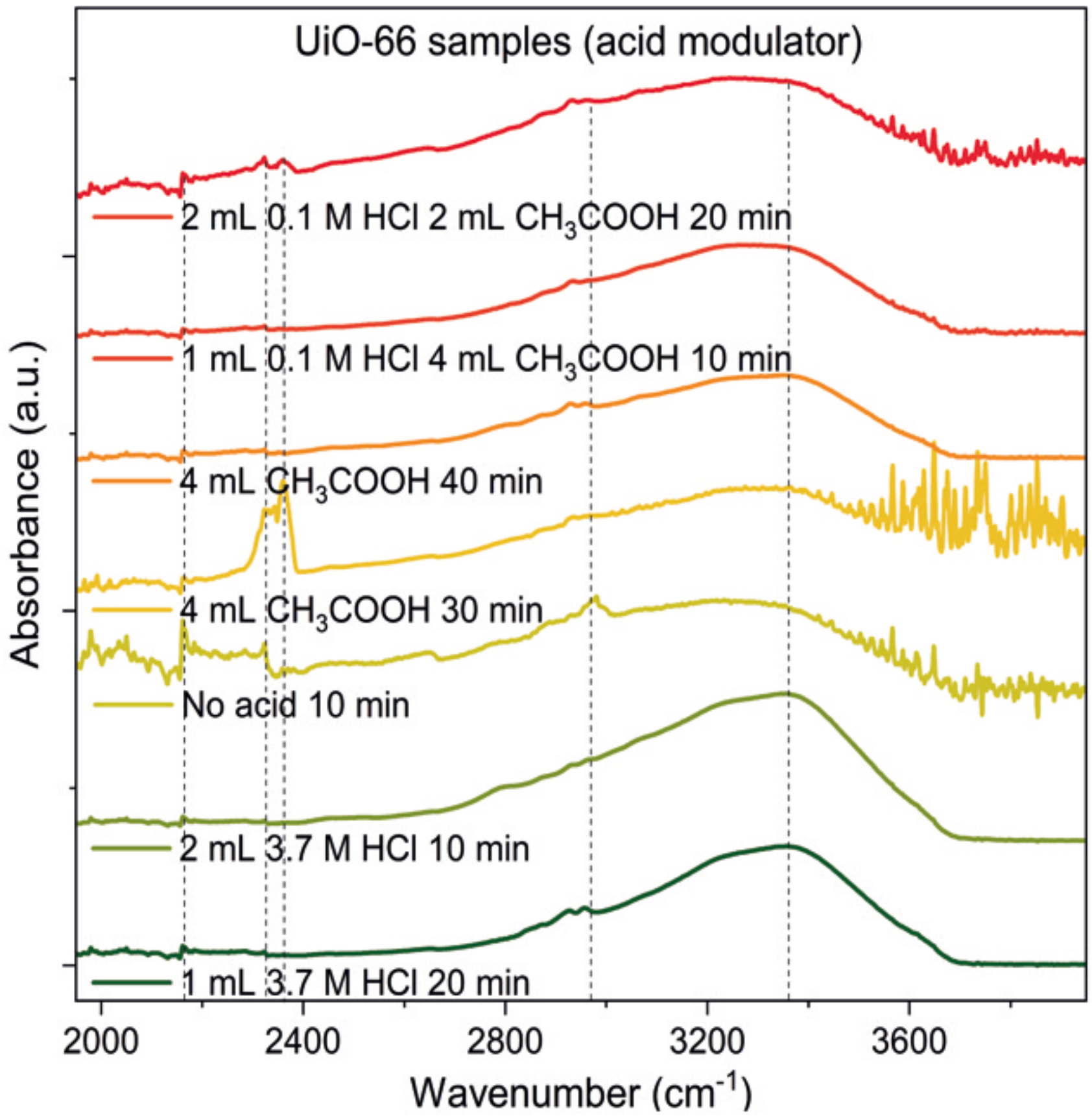

**Fig. S12** Attenuated Total Reflectance Fourier-Transform Infrared (ATR-FTIR) spectra of sonochemically-synthesised UiO-66 made with different modulation conditions.

## 2.3 Thermogravimetric Analysis (TGA)

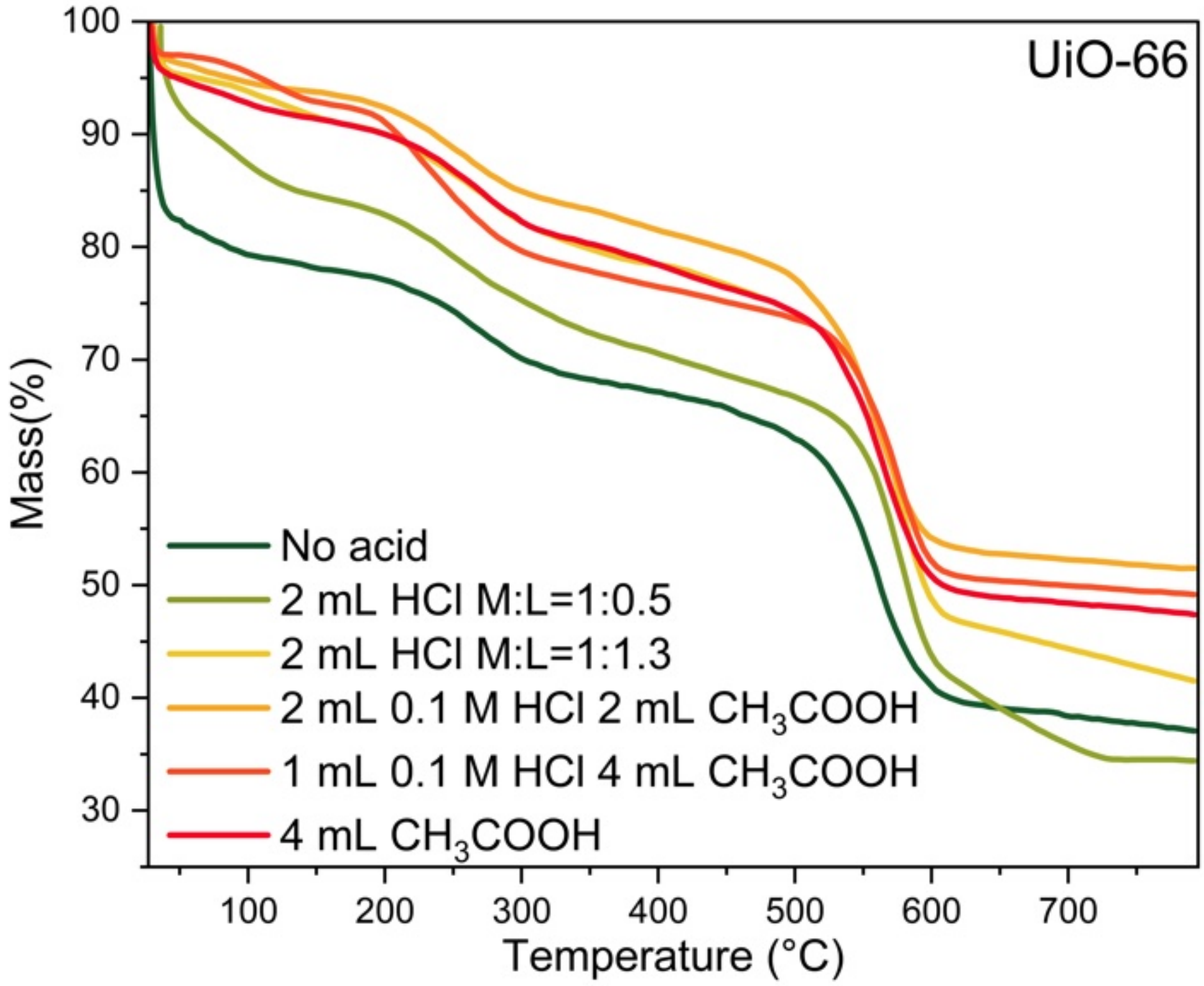

**Fig. 13** TGA curves of UiO-66 synthesised under different conditions.

**Estimation of the Guest Loading Weight Content (wt%) within the MOF Hosts from Thermogravimetric Analysis (TGA) Data**

The luminescent guest (LG) weight loading within the UiO MOF host was estimated from the difference in the final inorganic residue between pristine UiO-66 and RB+F@UiO-67 due to the limited convenience from external collaborators, giving 2.49 wt% LG. The molar loading was calculated according to

$$n_{LG}/n_{MOF} = (m_{LG}/m_{MOF}) \times (M_{MOF}/M_{LG})$$

, yielding 0.0888 mmol LG per mmol UiO-66. To estimate the corresponding loading in UiO-67, the same defect concentration was assumed for the isoreticular frameworks, and the loading was corrected by the molecular-weight ratio of UiO-67 to UiO-66, giving 0.113 mmol mmol$^{-1}$.

Assuming three accessible cages per UiO framework formula unit according to the UiO-66 and UiO-67 unit cell structures, the average guest occupancy was estimated to be 0.030 and 0.038 dye molecules per pore for UiO-66 and UiO-67, respectively.

# §3 Photophysical studies

## 3.1 Excitation and Emission Profiles

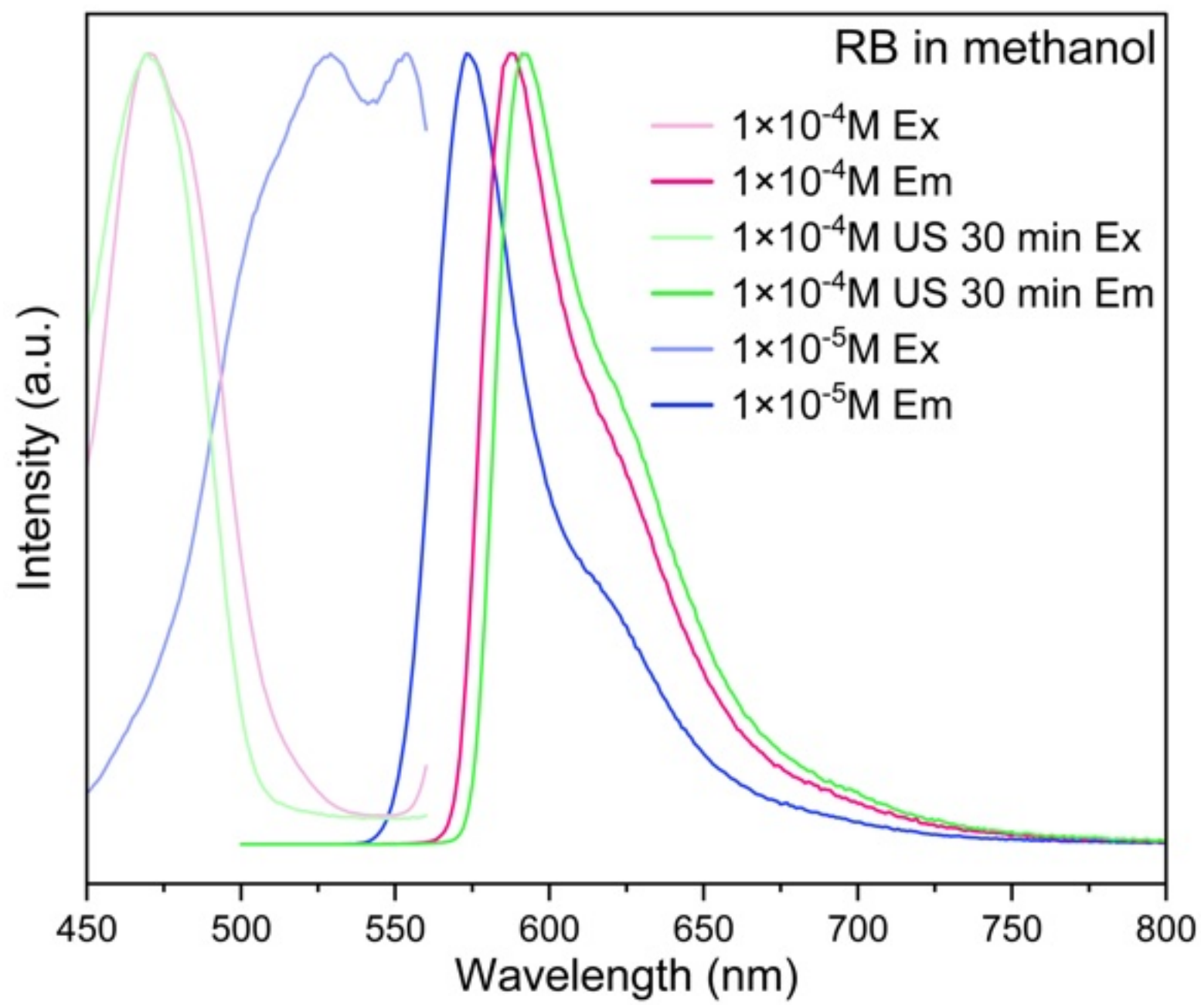

**Fig. S14(a)** Excitation and emission profiles of liquid RB in methanol.

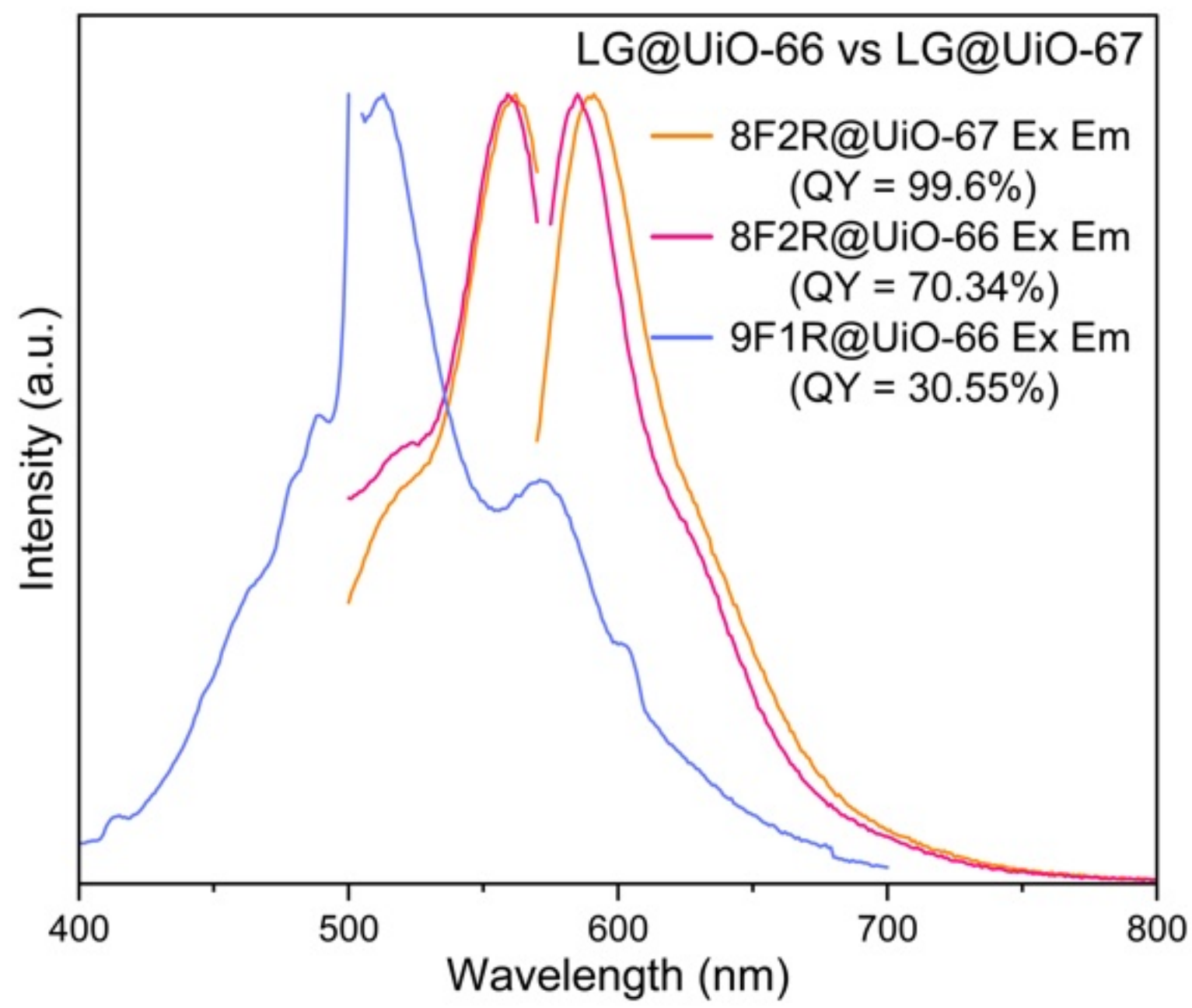

**Fig. S14(b)** Comparison between LG@UiO-66 and LG@UiO-67 in terms of Ex/Ex profiles and QY.

### 3.2 Excitation-Emission (Ex-Em) Maps

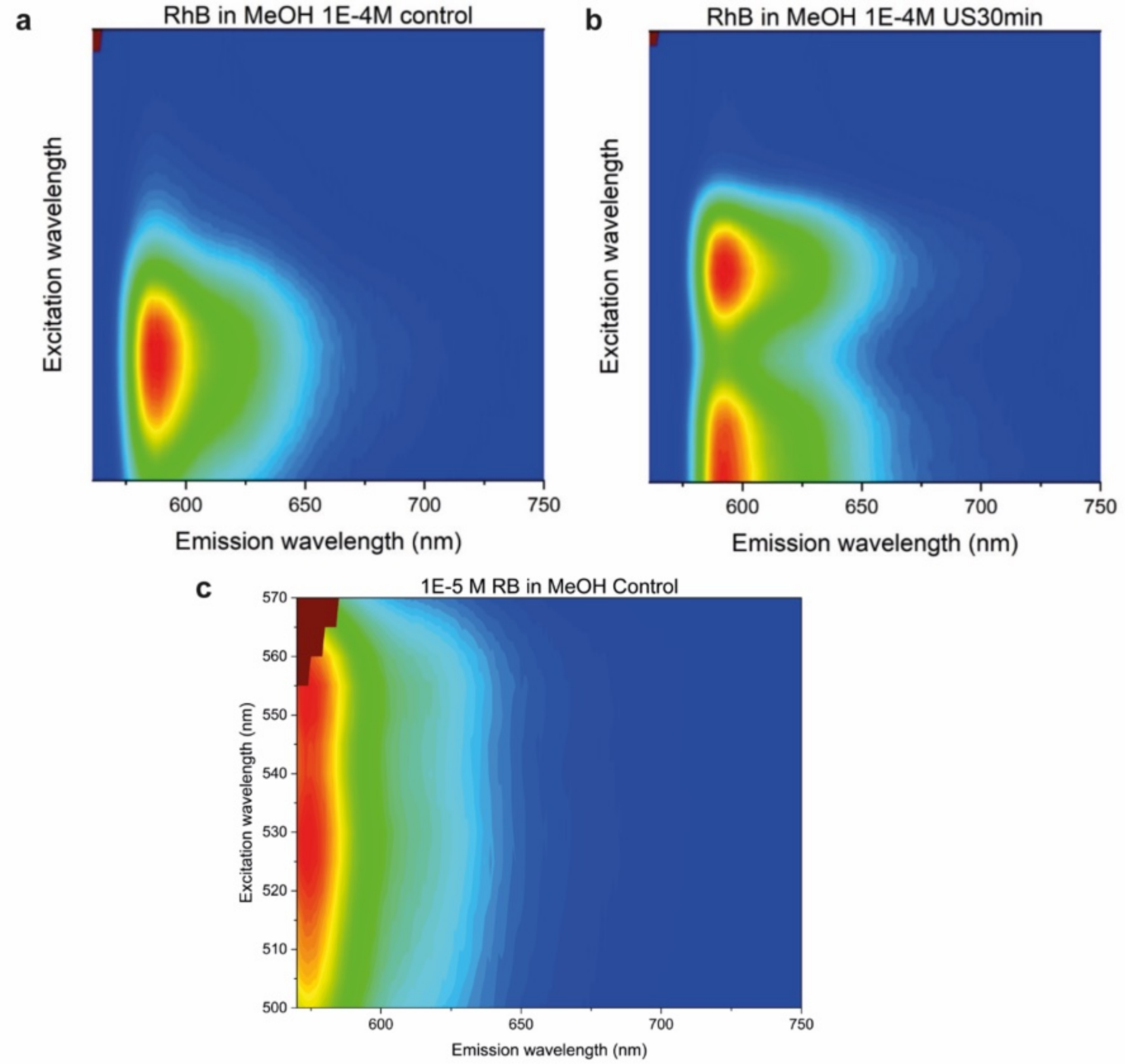

**Fig. S15** Ex-Em maps of **(a)** $1\times10^{-4}$ M RB (in methanol) solution without any sonication treatment. **(b)** $1\times10^{-4}$ M RB (in methanol) solution after 30-min ultrasonication. **(c)** $1\times10^{-5}$ M RB (in methanol) solution without any sonication treatment.

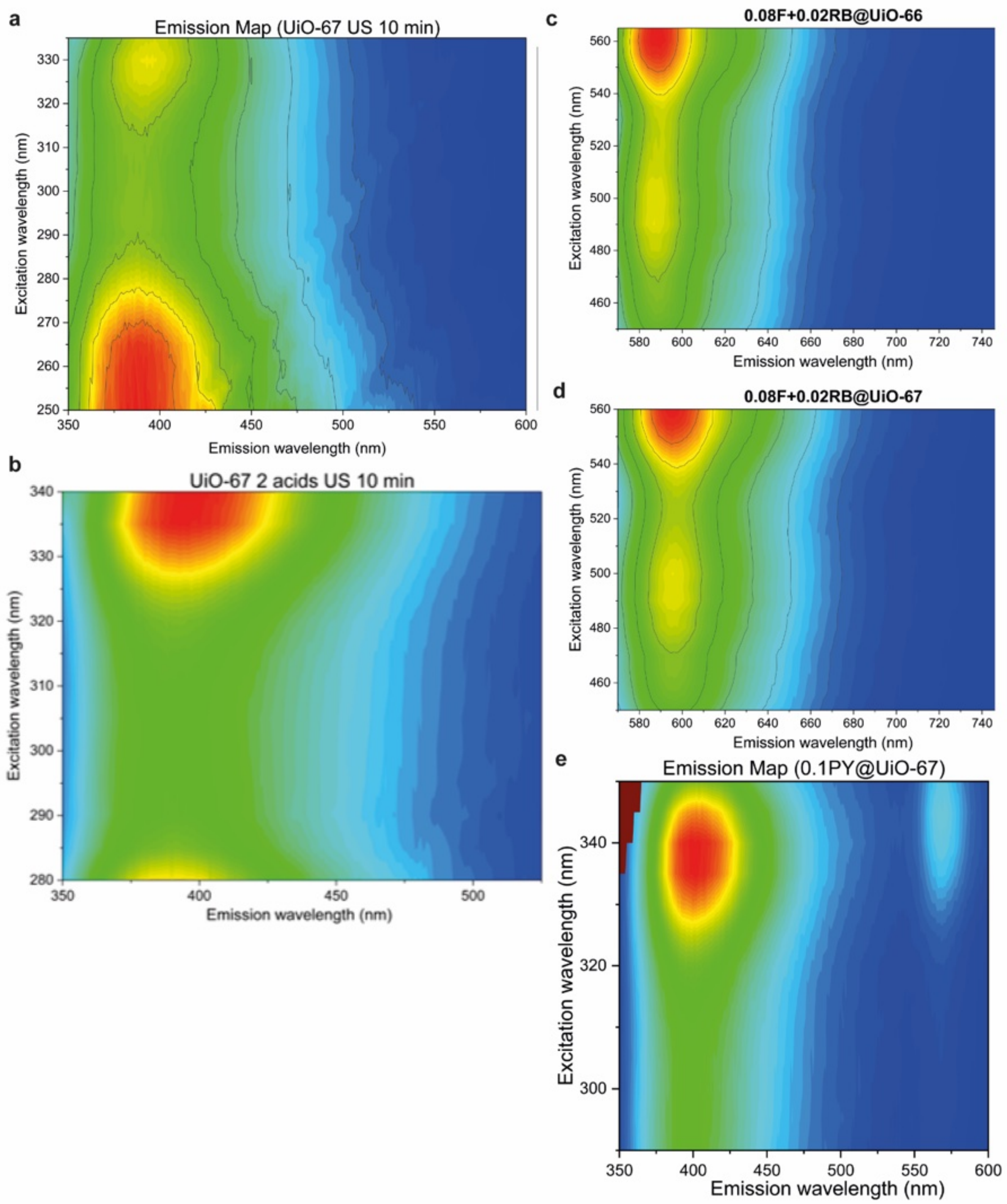

**Fig. S16** Ex-Em maps of **(a)** UiO-67 synthesised sonochemically in 10 min. **(b)** UiO-67 made sonochemically with 2-acid protocol. **(c)** 0.08F+0.02RB@UiO-66 synthesised with the 2-acid protocol. **(d)** 0.08F+0.02RB@UiO-67 synthesised with the 2-acid protocol. **(e)** 0.1PY@UiO-67 synthesised sonochemically in 2 h.

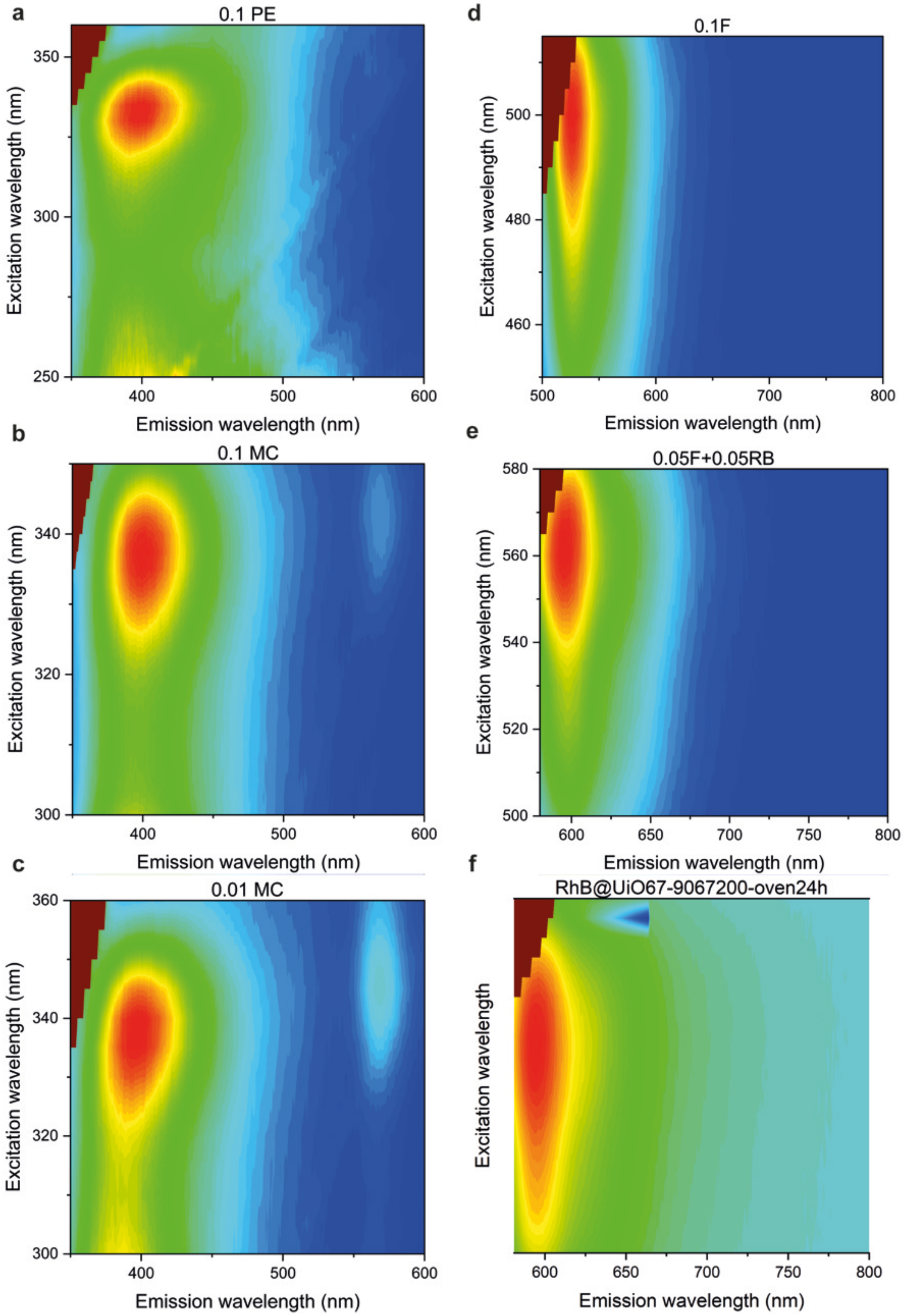
a
0.1 PE
Excitation wavelength (nm)
Emission wavelength (nm)
b
0.1 MC
Excitation wavelength (nm)
Emission wavelength (nm)
c
0.01 MC
Excitation wavelength (nm)
Emission wavelength (nm)
d
0.1F
Excitation wavelength (nm)
Emission wavelength (nm)
e
0.05F+0.05RB
Excitation wavelength (nm)
Emission wavelength (nm)
f
RhB@UiO67-9067200-oven24h
Excitation wavelength
Emission wavelength (nm)

**Fig. S17** **(a)** 0.1PE@UiO-67 synthesised sonochemically in 2 h. **(b)** 0.1MC@UiO-67 synthesised sonochemically in 2 h. **(c)** 0.01MC@UiO-67 synthesised sonochemically in 2 h. **(d)** 0.1F@UiO-67 synthesised sonochemically in 2 h. **(e)** 0.05F+0.05RB@UiO-67 synthesised sonochemically in 2 h. **(f)** RB@UiO-67 synthesised solvothermally in 24 h.

### 3.3 Photoluminescent Quantum yield

**Table. S1** Quantum yield (QY) of 0.08F+0.02RB@UiO-67 synthesised under different conditions.

| **8F2R@UiO-67** | No acid | 2 mL HCl | 4 mL HCl | solvothermal | 2-acid |
|---|---|---|---|---|---|
| **QY (%)** | 37.5 | 90.6 | 99.6 | 97.9 | 74.6 |

**Table. S2** QY of RB@UiO-67 synthesised with different guest loads or under different conditions.

| **RB@UiO-67** | **0.02, 2 h** | **0.1, 1 h** | **0.1, 2 h** | **0.5, 1 h** | **0.1, 2-acid, 2 h** |
|---|---|---|---|---|---|
| **QY (%)** | 33.0 | 30.9 | 59.2 | 7.8 | 66.9 |

**Table. S3** QY of 0.1F@UiO-67 made with different amounts of acid modulator.

| **0.1F@UiO-67** | **4 mL HCl** | **2 mL HCl** |
|---|---|---|
| **QY (%)** | 97.1 | 81.6 |

**Table. S4** QY of RB+F@UiO-67 made with different guest loadings.

| **Material** | **9F1R@UiO-66** | **8F2R@UiO-66** | **5F5R@UiO-67** |
|---|---|---|---|
| **QY (%)** | 30.6 | 70.9 | 92.8 |

**Table. S5** QY of blue emitters@UiO-67 synthesised with different guest loadings or under different conditions.

<table>
<tr><th>Guest (mmol) per 1 mmol UiO-67 added during synthesis QY (%)</th><th colspan="2">MC</th><th>PY</th><th>PE</th><th>MB</th><th>HC</th><th>C</th></tr>
<tr><td>10</td><td colspan="2">-</td><td>N/A</td><td>-</td><td>-</td><td>N/A</td><td>--</td></tr>
<tr><td>5</td><td colspan="2">-</td><td>-</td><td>-</td><td>-</td><td>6.3</td><td>-</td></tr>
<tr><td>1</td><td colspan="2">-</td><td>2.5</td><td>-</td><td>-</td><td>5.8</td><td>-</td></tr>
<tr><td rowspan="2">0.1</td><td>2 h</td><td>3 h</td><td rowspan="2">13.3</td><td rowspan="2">0.8</td><td rowspan="2">N/A</td><td rowspan="2">-</td><td rowspan="2">3.0</td></tr>
<tr><td>3.8</td><td>1.2</td></tr>
<tr><td>0.01</td><td colspan="2">2.1</td><td>0.3</td><td>-</td><td>-</td><td>-</td><td>-</td></tr>
<tr><td>0.001</td><td colspan="2">8.7</td><td>-</td><td>-</td><td>-</td><td>-</td><td>-</td></tr>
</table>

Notes:

MC = 7-methoxycoumarin

PY = Pyrene

MB = Methylene Blue

HC = 7-hydrocoumarin

C = Coumarin

'-' indicates incomplete measurement, while 'N/A' indicates negative measured value that can be interpreted as net absorption.

### 3.4 Fluorescence lifetime of blue emitters@UiO-67

**Table. S6** Fluorescence lifetime of blue emitters@UiO-67.

| **Guest (mmol) per 1 mmol UiO-67 added during synthesis** | **0.1 MC** | **0.1 C** | **0.1 PE** | **0.1 PY** |
|---|---|---|---|---|
| Max Em. (nm) | 405 | 415 | 410 | 425 |
| $\tau_1$ (ns) | 0.52 | 0.74 | 0.58 | 0.84 |
| $\tau_2$ (ns) | 2.16 | 2.48 | 2.11 | 2.56 |
| $\tau_3$ (ns) | 7.59 | 8.06 | 7.96 | 8.45 |
| $\chi^2$ | 1.113 | 1.023 | 1.149 | 1.173 |

## 3.5 CIE Colour Space

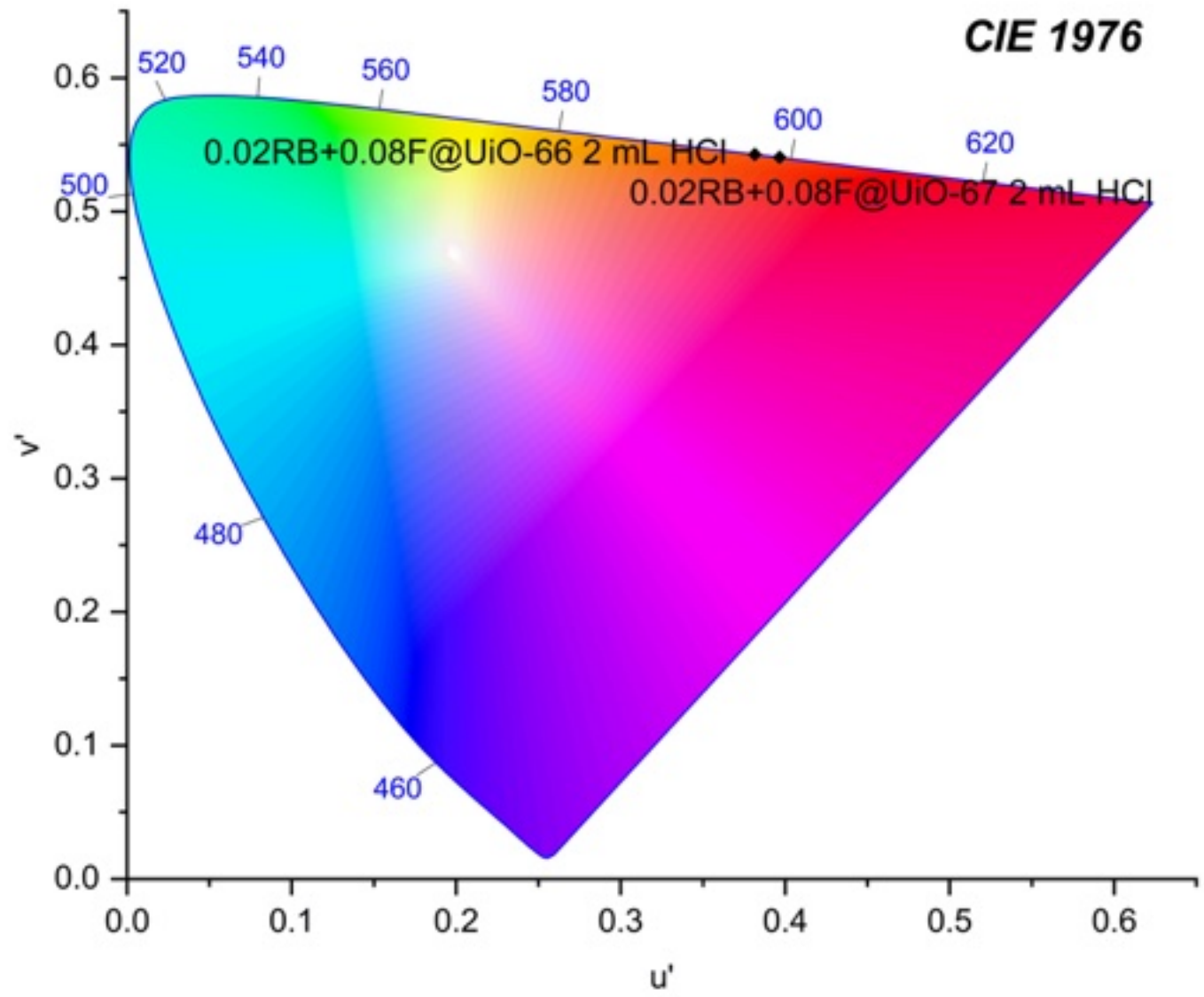

**Fig. S18** CIE 1976 chart of 8F2RB@UiO-66 and 8F2RB@UiO-67.

# §4 Miscellaneous

## 4.1 Solid-State Lighting (SSL) Device

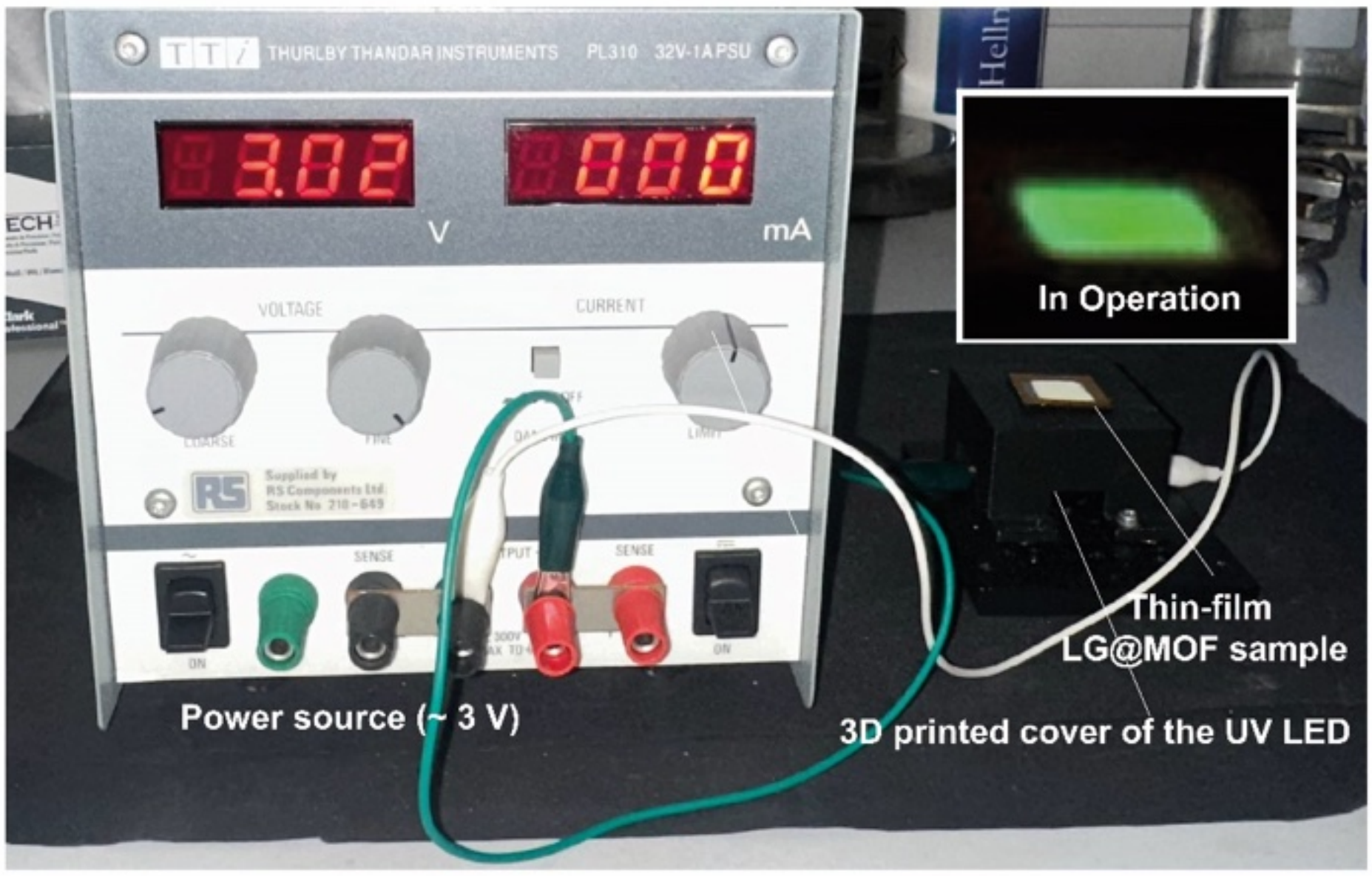

**Fig. S19** Photo of the SSL device tested under ambient light (left) and in the dark (right) under a UV LED excitation of 365 nm.

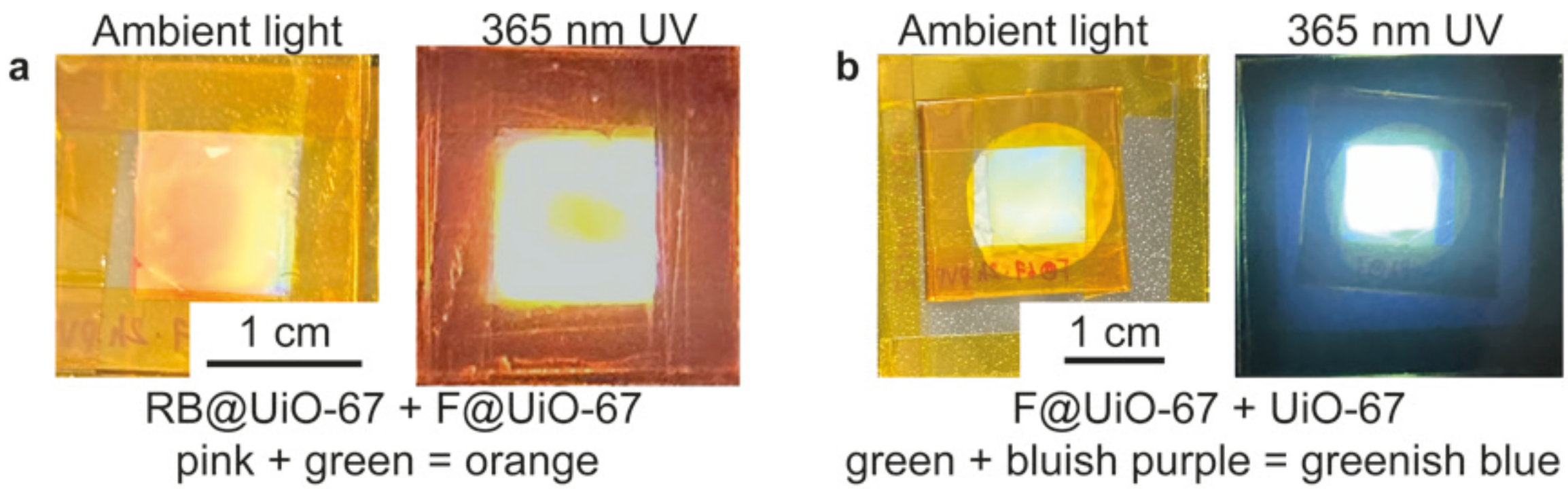

**Fig. S20** Light conversion device comprising layering of luminescent thin films. **(a)** RB@UiO-67 (bottom) + F@UiO-67 (top). **(b)** UiO-67 (bottom, 4 cm × 4 cm) + F@UiO-67 (top, 2 cm × 2 cm).

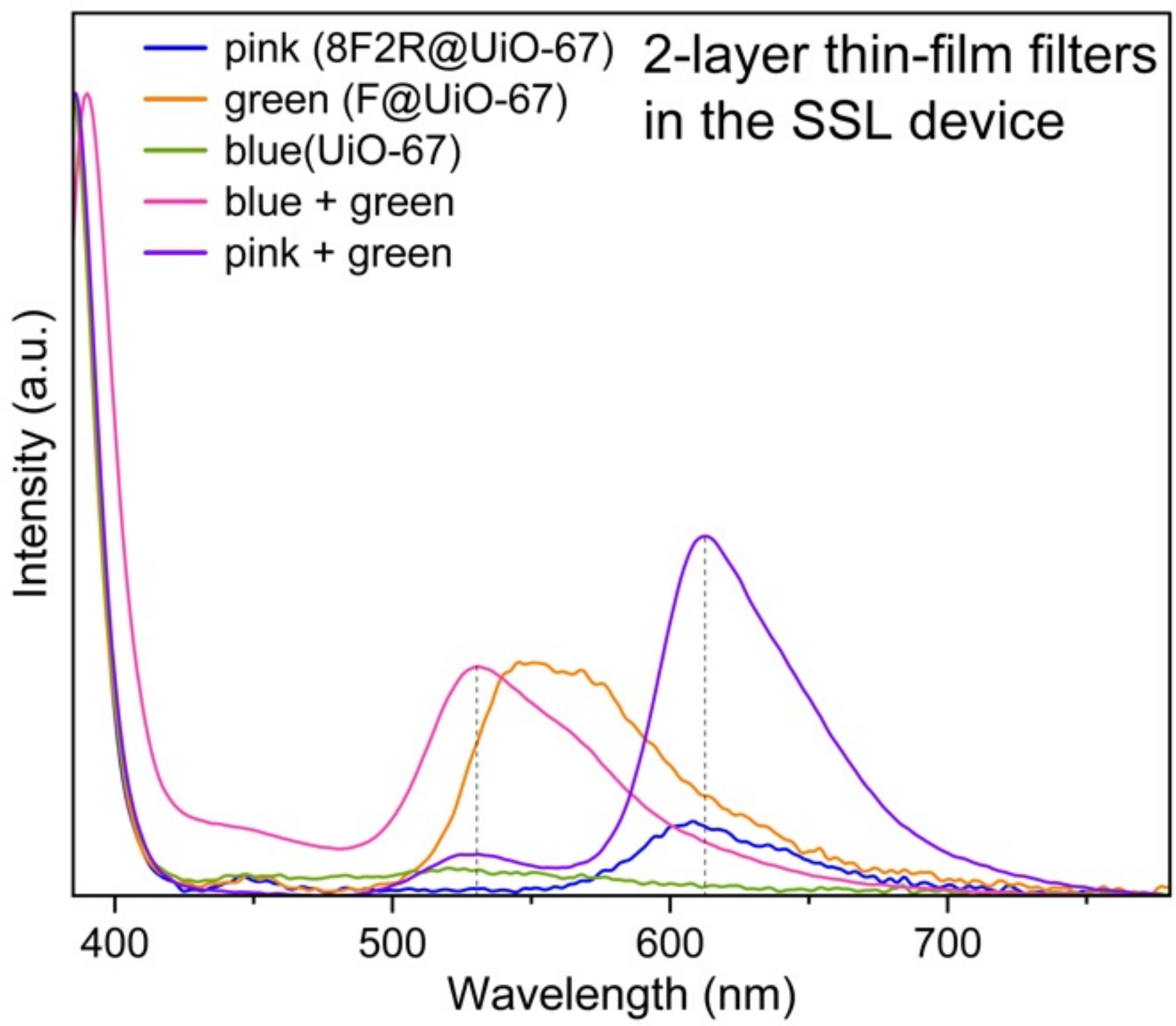

**Fig. S21** Emission profiles of the 2-layer thin film filters as in **Fig. S20** and their component single-layer filters measured with UPRTek 350N.

## 4.2 EDS Elemental Analysis in FESEM

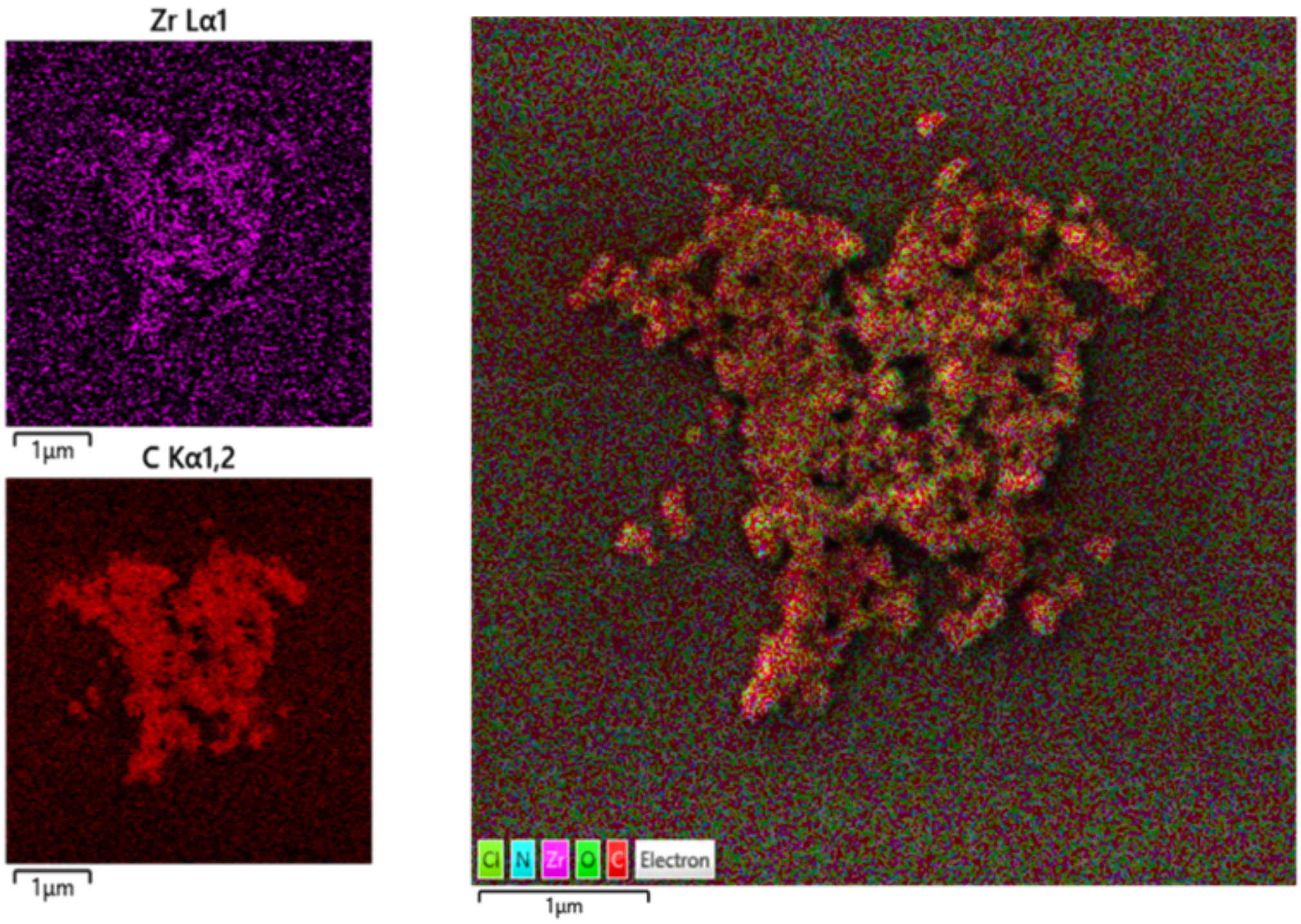

**Fig. S22** EDS maps of UiO-67 synthesised sonochemically without acid modulator.